\documentclass[]{jfm}

\usepackage{graphicx}
\usepackage{epstopdf, epsfig}

\usepackage{float}

\usepackage{amsmath}
\usepackage[colorlinks, citecolor=blue]{hyperref}
\usepackage[usenames,dvipsnames]{xcolor}
\usepackage{setspace}

\usepackage{siunitx}

\newcommand*\red{\textcolor{red}}
\newcommand{\revRev}[1]{\textcolor{black}{#1}}

\shorttitle{Drop impact on viscous liquid films}
\shortauthor{V. Sanjay, S. Lakshman,  P. Chantelot, J. H. Snoeijer and D. Lohse}

\title{Drop impact on viscous liquid films}
\author{
	Vatsal Sanjay\aff{1}
	\thanks{Contributed equally to this work.}
	\corresp{\email{vatsalsanjay@gmail.com}},
	Srinath Lakshman\aff{1} 
	\footnotemark[1]
	\corresp{\email{s.lakshman@utwente.nl}},
	Pierre Chantelot\aff{1}
	\footnotemark[1]
	\corresp{\email{p.r.a.chantelot@utwente.nl}},
	Jacco H. Snoeijer\aff{1}
	\corresp{\email{j.h.snoeijer@utwente.nl}},
	\and Detlef Lohse{\aff{1}$^{,}$\aff{2}}
	\corresp{\email{d.lohse@utwente.nl}}}

\affiliation{\aff{1}Physics of Fluids Group, Max Planck Center for Complex Fluid Dynamics, Department of Science and Technology, MESA+ Institute for Nanotechnology, and J. M. Burgers Centre for Fluid Dynamics, University of Twente, P. O. Box 217, 7500 AE Enschede, The Netherlands
	\aff{2}Max Planck Institute for Dynamics and Self-Organization, Am Fassberg 17, 37077 G\"{o}ttingen, Germany}

\newcommand{\Ohd}{\mathit{Oh}_\mathit{d}}
\newcommand{\Ohf}{\mathit{Oh}_\mathit{f}}

\newcommand{\Wen}{\mathit{We}_\mathit{d}}
\newcommand{\Bon}{\mathit{Bo}_\mathit{d}}

\begin{document}
	\maketitle
	\begin{abstract}
		When a liquid drop falls on a solid substrate, the air layer in between them delays the occurrence of liquid--solid contact. 
		For impacts on smooth substrates, the air film can even prevent wetting, allowing the drop to bounce off with dynamics identical to that observed for impacts on superamphiphobic materials. 
		In this article, we investigate similar bouncing phenomena, occurring on viscous liquid films, that mimic atomically smooth substrates, with the goal to probe their effective repellency.
		We elucidate the mechanisms associated to the bouncing to non-bouncing (floating) transition using experiments, simulations, and a minimal model that predicts the main characteristics of drop impact, the contact time, and the coefficient of restitution.
		In the case of highly viscous or very thin films, the impact dynamics is not affected by the presence of the viscous film. Within this substrate--independent limit, bouncing is suppressed once the drop viscosity exceeds a critical value as on superamphiphobic substrates. For thicker or less viscous films, both the drop and film properties influence the rebound dynamics and conspire to inhibit bouncing above a critical film thickness. This substrate--dependent regime also admits a limit, for low viscosity drops, in which the film properties alone determine the limits of repellency. 
	\end{abstract}
	
	
	\section{Introduction}
	\label{sec:Intro}
	
	Liquid drop impact on solids and liquids abound in nature \citep{Yarin2017} and are essential for several industrial applications, such as inkjet printing \citep{lohse2022fundamental} and criminal forensics \citep{smith2018influence}. Consequently, drop impact has garnered extensive attention \citep{rein1993phenomena, weiss1999single, thoroddsen2008high, yarin2006drop, josserand2016drop} ever since the seminal work of \citet{worthington1877xxviii, worthington2019study}. Impacts can result in either contact or levitation outcomes, depending on whether the air layer trapped between the drop and the substrate drains completely during impact. 
	
	For low impact velocities, the buildup of the lubrication pressure in the draining air layer prevents the drop from contacting with the underlying surface, leading to drop bouncing/floating on this layer \citep{reynolds1881floating,davis1989lubrication, yiantsios1990buoyancy, yiantsios1991close, smith2003air, van2012direct}. Drops that bounce/float in such a scenario are realized in several configurations, for example on solid surfaces \citep{kolinski2014drops, de2015wettability}, liquid films \citep{pan2007dynamics, hao2015superhydrophobic, tang2018bouncing, tang2019bouncing}, stationary \citep{rodriguez1985some, klyuzhin2010persisting, wu2020small} or vibrating liquid pools \citep{couder2005bouncing, couder2005walking}, or even soap films \citep{gilet2009fluid}. 
	Interfacial processes such as Marangoni flow \citep{geri2017thermal} or the generation of vapor below a drop deposited on a superheated substrate \citep[the Leidenfrost effect where the liquid levitates on a cushion of its own vapor,][]{leidenfrost1756,quere2013leidenfrost, chantelot2021leidenfrost} can further stabilize the sandwiched air/vapor layer to facilitate levitation, even for the dynamic case of drop impact \citep{chandra1991collision,tran2012drop,shirota2016dynamic}. Drops can also defy gravity and levitate thanks to the so--called inverse Leidenfrost effect \citep{adda2016inverse, gauthier2019self}, or electromagnetic forces \citep{pal2017control, singh2018levitation}. 
	
	At higher impact velocities, the air layer ruptures, leading to contact. The rupture occurs due to a strong van der Waals attractive force between the droplet and the solid or liquid substrate, which comes into play as the thickness of the air layer reduces below the order of $10 - 100\,\si{\nano\meter}$ \citep[see appendix~\ref{App:FilmRupture}, and][]{charles1960coalescence, SprittlesPhysRevLett.124.084501, zhang2021thin}. Additionally, surface asperities that are of the order of the minimum gas layer thickness can also cause rupture, binding the drop to the surface \citep{thoroddsen2003air, kolinski2014drops, li_vakarelski_thoroddsen_2015}. 
	
	\revRev{In this work, we focus on levitation outcomes which can be classified as either repellent (bouncing drops) or non-bouncing (non-repellent/floating) behaviors.
		We note that non-repellent scenarios ultimately lead to coalescence, a phenomenon we do not investigate here and that occurs on a time scale much larger than that of impact  \citep{lo2017mechanism, duchemin2020dimple}.}
	We perform experiments and direct numerical simulations (DNS) to investigate drop rebound on viscous liquid films.
	In the limit of thin enough viscous coatings, the substrate mimics an atomically smooth solid and displays a superamphiphobic-like repellent behavior \citep{hao2015superhydrophobic, lo2017mechanism}. This substrate--independent bouncing \citep{gilet2012droplets, pack2017failure, lakshman2021deformation} can be compared with that observed on superhydrophobic substrates, where the apparent contact time is given by the oscillation time of a drop \citep{rayleigh1879capillary}, owing to the drop impact-oscillation analogy \citep{richard2002contact}.
	As a result, such an impacting drop can be modeled using a quasi-ideal spring, whose stiffness is given by the surface tension coefficient \citep{okumura2003water}.
	Unlike ideal Rayleigh oscillations, the collisions are partially inelastic due to viscous dissipation \citep{prosperetti1977viscous}. 
	When the drop viscosity increases and viscous dissipation becomes significant, this spring couples with a linear damper whose strength is proportional to the drop's viscosity \citep[see appendix~\ref{app:SubstrateIndependentBouncing}, and][]{jha2020viscous}. The adoption of such a spring-mass-damper system has led to several successful predictions of the drop impact dynamics in a variety of configurations such as viscous bouncing \citep{molavcek2012quasi, jha2020viscous}, spontaneous levitation \citep{schutzius2015spontaneous}, fast bouncing \citep{chantelot2018drop}, and walking drops \citep{terwagne2013role}. 
	
	In the opposing limit of thick liquid films (pools), drops can also bounce/float \citep{reynolds1881floating, jayaratne1964coalescence}. However, unlike solids and very thin films, these pools deform on impact and can absorb a part of the impact kinetic energy in the form of (i) surface energy owing to interfacial deformation, (ii) internal kinetic energy, and (iii) viscous dissipation \citep{galeano2021capillary}. Consequently, the impact outcomes also include a substrate--dependent regime which culminates in the transition from bouncing to \revRev{non-bouncing (floating)}. In the latter case, the drop cannot take off, resulting in the liquid surface loosing its repellent property. \citet{hao2015superhydrophobic} studied this transition from the substrate--independent to substrate--dependent bouncing for water drops, and reported that the critical film thickness marking this transition depends on the film viscosity and the impact velocity of the drops.
	
	In the present work, we elucidate how the thickness and viscosity of liquid coatings influence the rebound characteristics of an impacting drop, culminating in the loss of repellency\revRev{: the transition from bouncing to non-bouncing (floating).}
	We disentangle how the initial kinetic energy of an impacting drop can be absorbed through dissipation and energy transfers in the drop and the liquid film. 
	
	The paper is organized as follows: \S~\ref{sec:Methods} describes the experimental and numerical methods. In \S~\ref{sec:Phenomenology}, we discuss the phenomenology of the drop impact dynamics on viscous liquid films. 
	Guided by our experimental and numerical observations, we develop a phenomenological model in \S~\ref{sec:Phenomenological_model}, extending on the spring-mass-damper analogy by considering the liquid coating as an additional source of dissipation. In \S~\ref{sec:influence}, we test the validity and applicability of this model by comparing the predicted values of the coefficient of restitution with our observations when both the drop and film properties are varied. 
	We also delineate the various regimes observed in this work by measuring the critical film thicknesses at which the substrate--independent to substrate--dependent and bouncing to non-bouncing (floating) transitions occur and compare their observed values with the model predictions. Further, \S~\ref{sec:EnergySection} investigates the cases where our phenomenological model fails to predict the observed dynamics and gives alternate explanations. The paper ends with a conclusions and an outlook (\S~\ref{sec:Conclusions and Outlook}).
	
	\begin{figure}
		\centering
		\includegraphics[width=0.9\textwidth]{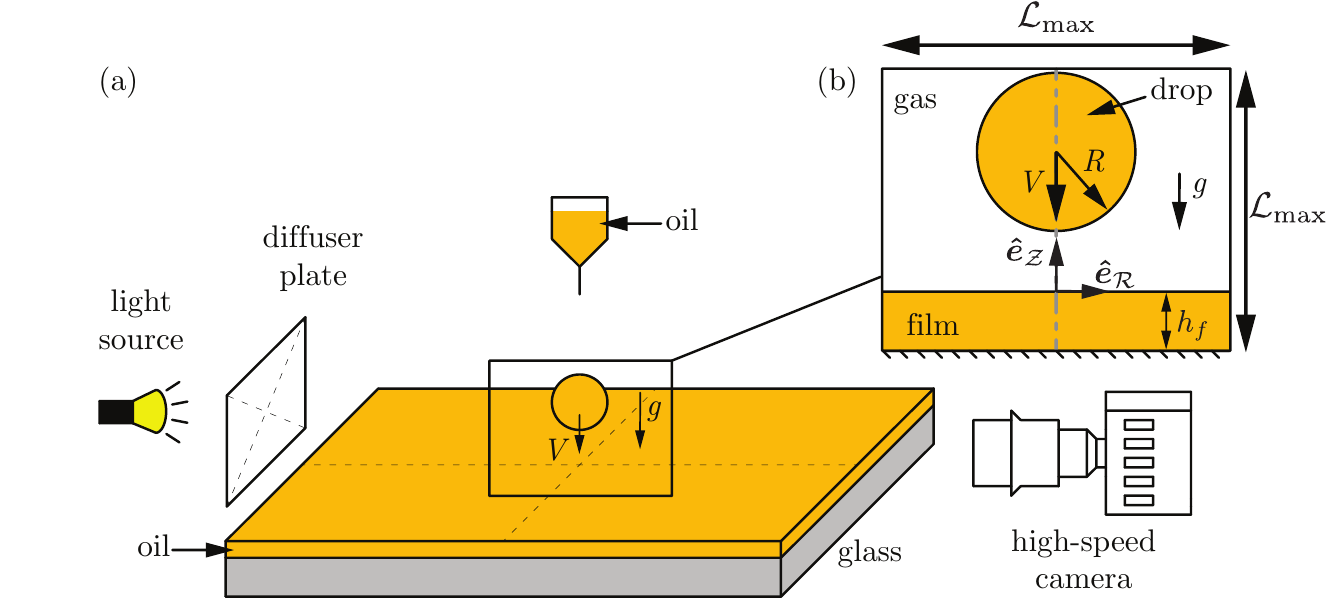}
		\caption{(a) Schematic (not to scale) of the experimental setup. (b) Side view visualization of the drop impact process as viewed using the high-speed camera. The inset also shows the axi-symmetric domain used in the direct numerical simulations and defines the used symbols. \revRev{The domain boundaries are chosen to be far enough not to influence the drop impact process. Furthermore, we ensure that the waves formed on the film are not reflected back from these boundaries. Consequently, for $Oh_f < 0.1$, $\mathcal{L}_{\text{max}} \gg 8R$. On the other hand, $Oh_f > 0.1$ and waves on the film are damped, we choose $\mathcal{L}_{\text{max}} = 8R$.}}
		\label{fig:schematic}
	\end{figure}
	
	\section{Methods}
	\label{sec:Methods} 
	\subsection{Experimental details} \label{subsec:exp methods}
	\begin{table}
		\begin{center}
			\def~{\hphantom{0}}
			\begin{tabular}{lccc}
				Silicone oil & $\rho$ & $\eta$  & $\gamma$\\
				&(kg/m$^{3}$) &(mPa.s)& (mN/m)\\[3pt]
				SE 1 & 818 & 0.8 & 17 \\
				AK 5 & 920 & 4.6 & 19 \\
				AK 10 & 930 & 9.3 & 20 \\
				AK 20 & 950 & 19 & 21 \\
				AK 35 & 960 & 34 & 21 \\
				AK 50 & 960 & 48 & 21 \\
				AK 100 & 960 & 96 & 21 \\
			\end{tabular}
			\caption{Properties of the liquids used in the experiments. \revRev{$\rho$ and $\eta$ are the density of the liquids and $\gamma$ denotes the liquid-air surface tension coefficient. Throughout the manuscript, the subscripts $d$, $f$, and $s$ represent drop, film, and surrounding, respectively.} The silicone oil manufacturers are Shin Etsu (SE) and Wacker Chemie AG (AK).}
			\label{tab:table00}
		\end{center}
	\end{table}
	
	Our experiments, whose setup is sketched in figure \ref{fig:schematic}, consist of silicone oil droplets with radius $R$, density $\rho_d$, and viscosity $\eta_d$, impacting on silicone oil films with thickness $h_f$, density $\rho_f$, and viscosity $\eta_f$. 
	We choose silicone oil as a working fluid as its viscosity can be varied over a wide range, here from $0.8\,\si{\milli\pascal}.\si{\second}$ to $96\,\si{\milli\pascal}.\si{\second}$, while keeping its density and surface tension coefficient $\gamma$ nearly constant, as evidenced in table~\ref{tab:table00}. 
	Droplets with radius $R = 1.0 \pm 0.1\,\si{\milli\meter}$ are released from a calibrated needle whose height can be varied to adjust the impact velocity $V$ from $0.1\,\si{\meter}/\si{\second}$ to $0.5\,\si{\meter}/\si{\second}$. 
	The rupture of the air layer, that mediates the interaction between the drop and the film, determines the upper bound of the bouncing regime.
	\revRev{This rupture} sets the \revRev{critical} impact velocity, expressed as the Weber number ({\it i.e.}, the ratio of inertial to capillary stresses) $\Wen \equiv \rho_d R V^{2} /\gamma \lesssim \mathcal{O}\left(10\right)$, \revRev{above which coalescence between the miscible drop and film occurs} \citep[see appendix~\ref{App:FilmRupture} and][]{SprittlesPhysRevLett.124.084501, sharma2021regimes}. 
	We further fix the impact velocity at $V = 0.3 \pm 0.03\,\si{\meter}/\si{\second}$, corresponding to $\Wen = 4\pm1$, and focus on the influence of the material properties of the drop and the film on the impact process (see \S~\ref{subsec:numerical methods}). Indeed, this process is fairly independent of $\Wen$ in the narrow range of $\Wen$ in which bouncing occurs without air layer rupture (see appendices~\ref{App:FilmRupture} and~\ref{sec:weber_influence}).  
	
	Films of controlled thickness, varying from $0.01\,\si{\milli\meter}$ to $1\,\si{\milli\meter}$, are prepared by spincoating the liquid for $h_f < 0.03\,\si{\milli\meter}$, or by depositing a known volume of silicone oil on a glass slide and allowing it to spread when $h_f > 0.03\,\si{\milli\meter}$.  We measure the thickness of spincoated films using reflectometry \citep{reizman1965optical}, with an uncertainty of $\pm 0.1\,\si{\micro\meter}$, while the thicker films obtained from the deposition method are characterized from side-view imaging, using a procedure detailed in appendix~\ref{sec:measuring_film_thickness}, with an uncertainty of $\pm 30\,\si{\micro\meter}$. We record the impact dynamics using high-speed side-view imaging at 10,000 frames per second (Photron UX100).
	
	\subsection{Governing equations \& Numerical framework}
	\label{subsec:numerical methods}
	This section describes the direct numerical simulation (DNS) framework used to study the drop impact process with the free software program, \emph{Basilisk C} \citep{popinet-basilisk}, using the volume of fluid method (VoF, equation~\eqref{Eqn::Vof2}) for tracking the interface \citep{tryggvason2011direct}. In this work, we have three fluids, namely, the drop, film, and air, denoted by ($d$), ($f$), and ($a$), respectively (figure~\ref{fig:schematic}). In order to track the three fluids and enforce non-coalescence between the drop and the film, we use two VoF tracer fields, $\Psi_1, \Psi_2$ \citep{ramirezsoto-2020-sciadv},
	
	\begin{equation}
		\label{Eqn::Vof2}
		\left(\frac{\partial}{\partial t} + \boldsymbol{v\cdot}\nabla\right)\{\Psi_1, \Psi_2\} = 0,
	\end{equation}
	
	\noindent where $\boldsymbol{v}$ is the velocity field. 
	The use of two VoF fields, followed by interface reconstruction and implicit tagging of the ambient medium (air tracer, $\Psi_a = 1 - \Psi_1 - \Psi_2$), ensures that the two tracers never overlap \citep{ramirezsoto-2020-sciadv, naru2021numerical}. As a result, there is always a thin air layer between the drop and the film. Our continuum-based simulations are thus not sufficient to predict the coalescence of interfaces \citep{SprittlesPhysRevLett.124.084501},
	and we obtain the bounds of the non-coalescence regime, which sets the maximal Weber number probed in our simulations, from experiments (see appendix \ref{App:FilmRupture} for details).
	
	We use adaptive mesh refinement (AMR) to resolve the length scales pertinent to capture the bouncing process, \emph{i.e.} the flow inside the drop and the liquid coating.
	The adaption is based on minimizing the error estimated using the wavelet algorithm \citep{popinet-2015-jcp} in the volume of fluid tracers, interfacial curvatures, velocity field, vorticity field and rate of viscous dissipation with tolerances of $10^{-3}$, $10^{-4}$, $10^{-2}$, $10^{-2}$, and $10^{-3}$, respectively  \citep{basiliskVatsalDropFilm}. 
	We ensure that at least $15$--$20$ grid cells are present across the minimum liquid film thickness ($\Gamma = h_f/R = 0.01$) studied in this work to resolve the velocity gradients in the film \citep{josserand2016droplet, ling2017spray}. 
	The minimum thickness of the air layer is of the order of the minimum grid size $\Delta = R/2048$. We further note that this thickness can be larger than this minimum owing to flow characteristics. For example, the shear stress balance across an interface with a high viscosity ratio delays the drainage of the air layer \citep{zhang_ni_magnaudet_2021}.  
	
	For an incompressible flow, the mass conservation requires the velocity field to be divergence-free
	
	\begin{align}
		\boldsymbol{\nabla\cdot v} = 0.
	\end{align}  
	
	\noindent Furthermore, the momentum conservation  reads (tildes denote dimensionless quantities)
	
	\begin{align}
		\label{Eqn::NS}
		\left(\frac{\partial}{\partial \tilde{t}} + \boldsymbol{\tilde{v}\cdot\tilde{\nabla}}\right)\boldsymbol{\tilde{v}} &= \frac{1}{\tilde{\rho}}\left(-\boldsymbol{\tilde{\nabla}} p + \boldsymbol{\tilde{\nabla}\cdot}\left(2Oh\boldsymbol{\tilde{\mathcal{D}}}\right)\right) - Bo\,\boldsymbol{\hat{e}}_{\boldsymbol{\mathcal{Z}}} + \boldsymbol{\tilde{f}}_\gamma,
	\end{align}
	
	\noindent where the coordinate dimensions, velocity field $\boldsymbol{v}$, and pressure $p$ are normalized using the drop radius $R$, inertio-capillary velocity scale $v_\gamma = \sqrt{\gamma/\rho_d R}$ and capillary pressure $p_\gamma = \gamma/R$, respectively. The bracketed term on the left hand side of equation~\eqref{Eqn::NS} is the material derivative. On the right hand side, $\boldsymbol{\hat{e}}_{\boldsymbol{\mathcal{Z}}}$ is a unit vector in the vertically upward direction (see figure~\ref{fig:schematic}b) and the deformation tensor, $\boldsymbol{\tilde{\mathcal{D}}}$ is the symmetric part of the velocity gradient tensor $\left(\boldsymbol{\tilde{\mathcal{D}}} = \left(\boldsymbol{\tilde{\nabla}\tilde{v}} + \left(\boldsymbol{\tilde{\nabla}\tilde{v}}\right)^{\text{T}}\right)/2\right)$. Further, we employ the one-fluid approximation \citep{tryggvason2011direct} to solve these equations whereby the material properties (such as dimensionless density $\tilde{\rho} = \rho/\rho_d$ and dimensionless viscosity $Oh$) change depending on which fluid is present at a given spatial location (equations~\eqref{Eqn::density}--\eqref{Eqn::Oh})
	
	\begin{align}
		\label{Eqn::density}
		\tilde{\rho} &= \Psi_1 + \Psi_2\frac{\rho_{f}}{\rho_{d}} + \left(1-\Psi_1-\Psi_2\right)\frac{\rho_{a}}{\rho_{d}}\\
		\label{Eqn::Oh}
		Oh &= \Psi_1Oh_d + \Psi_2Oh_f + \left(1-\Psi_1-\Psi_2\right)Oh_a
	\end{align}
	
	\noindent where, the Ohnesorge number $Oh$ is the ratio between the inertio-capillary to the visco-capillary time scales. It is defined for all the three phases, namely, the drop, the film, and the air (ambient), and are given by
	
	\begin{align}
		&Oh_d = \frac{\eta_d}{\sqrt{\rho_d\gamma R}},\\
		&Oh_f = \frac{\eta_f}{\sqrt{\rho_d\gamma R}},\,\text{and}\\
		&Oh_a = \frac{\eta_a}{\sqrt{\rho_d\gamma R}},
	\end{align}
	
	\noindent respectively. Here, $\eta_d$, $\eta_f$, and $\eta_a$ are the viscosity of the drop, film and air (ambient), respectively. Furthermore, $\rho_f/\rho_d$ and $\rho_a/\rho_d$ are the film-drop and air-drop density ratios. For simplification, we use $\rho_f/\rho_d = 1$ (also see table~\ref{tab:table00}). In order to keep the surrounding medium as air, $\rho_a/\rho_d$ and $Oh_a$ are fixed at $10^{-3}$ and $10^{-5}$, respectively. We also fix the Bond number (ratio of the gravitational to the capillary pressure) given by
	
	\begin{align}
		\Bon = \frac{\rho_dgR^2}{\gamma}
	\end{align}
	
	\noindent at $0.5$ during this study. The initial condition (figure~\ref{fig:schematic}b) is given by the normalized impact velocity, $\tilde{V} = \sqrt{\Wen}$. 
	
	Lastly, a singular body force $\boldsymbol{\tilde{f}}_\gamma$ is applied at the interfaces to respect the dynamic boundary condition across them. The approximate forms of these forces follow from \citet{brackbill1992continuum, prosperetti2009computational, tryggvason2011direct} as
	
	\begin{align}\label{Eqn::SurfaceTension}
		\boldsymbol{\tilde{f}}_\gamma \approx  \tilde{\kappa}_1\boldsymbol{\tilde{\nabla}}\Psi_1 +  \tilde{\kappa}_2\boldsymbol{\tilde{\nabla}}\Psi_2.
	\end{align}
	
	\noindent Here, $\kappa_1$ and $\kappa_2$ are the curvatures associated with $\Psi_1$ and $\Psi_2$, respectively, calculated using the height-function method.  During the simulations, the maximum time-step needs to be set less than the oscillation period of the smallest wavelength capillary wave as the surface-tension scheme is explicit in time \citep{popinet2009accurate, basiliskPopinet2}.
	
	Figure~\ref{fig:schematic}(b) represents the axi-symmetric computational domain. A tangential stress-free and non-penetrable boundary condition is applied on each of the domain boundaries. The pressure gradient is also set to zero at these boundaries. Furthermore, \revRev{the domain boundaries are chosen to be far enough not to influence the drop impact process. When $Oh_f > 0.1$ and waves on the film are damped, we choose $\mathcal{L}_{\text{max}} = 8R$.} 
	The cases with low $\Ohf$ require extra attention due to the train of \revRev{surface} waves formed post-impact as these waves can reflect back from the side--walls \revRev{(here, we choose $\mathcal{L}_{\text{max}} \gg 8R$)}.
	
	\section{Phenomenology of the impact events} \label{sec:Phenomenology}
	
	\begin{figure}
		\centering
		\includegraphics[width=\textwidth]{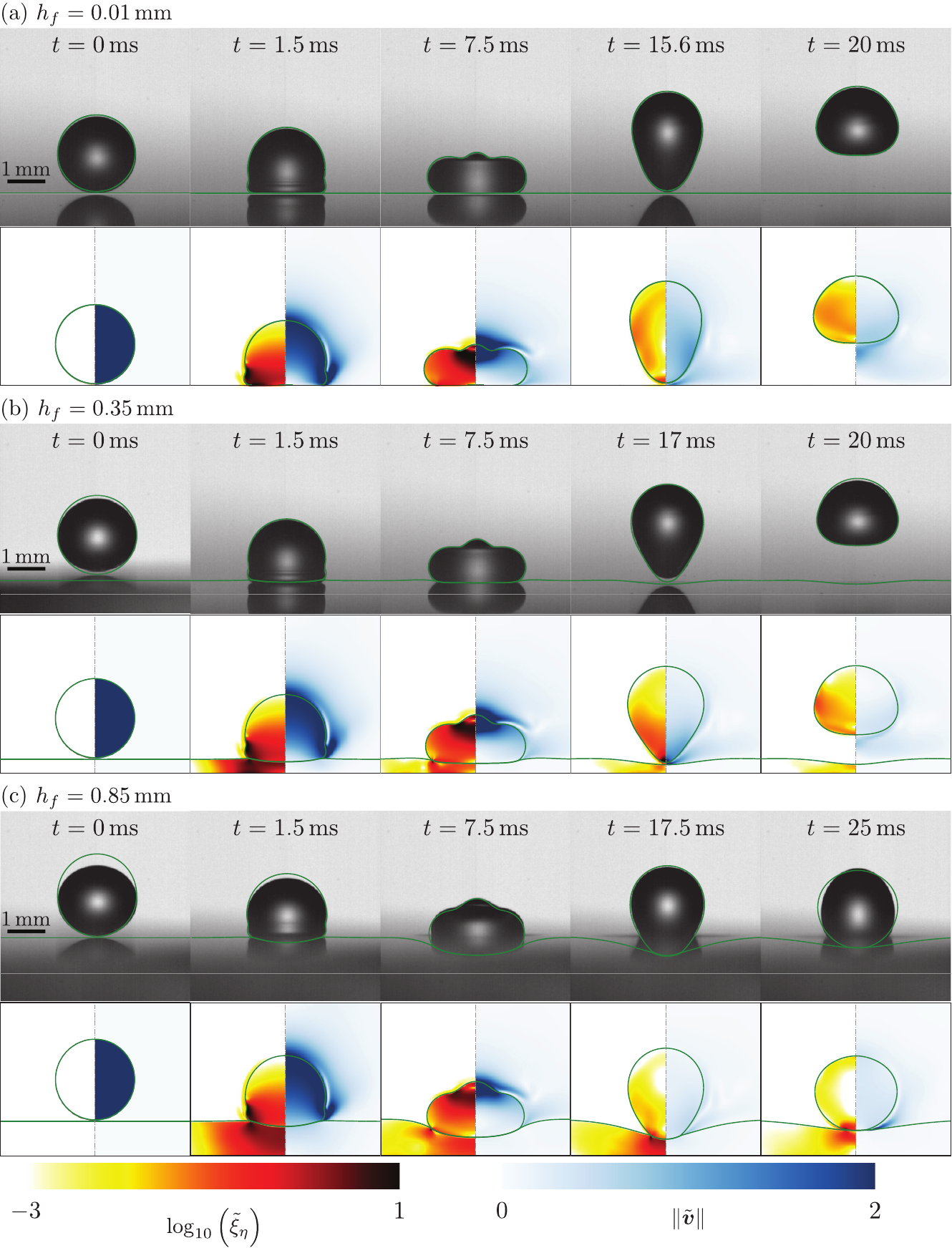}
		\caption{Effect of the film thickness on the drop impact process: comparison of the experimental and DNS snapshots of the impact process on films with $h_f = $ (a) $0.01\,\si{\milli\meter}$, (b) $0.35\,\si{\milli\meter}$, and (c) $0.85\,\si{\milli\meter}$. In each panel, the top row contains the experimental images with (green) interface outline from DNS, and the bottom row contains numerical snapshots showing the dimensionless rate of viscous dissipation per unit volume ($\tilde{\xi}_\eta = 2Oh\left(\boldsymbol{\tilde{\mathcal{D}}:\tilde{\mathcal{D}}}\right)$) on the left and the magnitude of dimensionless velocity field ($\boldsymbol{\tilde{v}}$) on the right. We show $\tilde{\xi}_\eta$ on a $\log_{\text{10}}$ scale to identify regions of maximum dissipation (marked with black for $\tilde{\xi}_\eta \ge 10$). For all cases in this figure, $R = 1\,\si{\milli\meter}$, $V = 0.3\,\si{\meter}/\si{\second}$, $\eta_{d} = 4.6\,\si{\milli\pascal}.\si{\second}$ and $\eta_{f} = 96\,\si{\milli\pascal}.\si{\second}$, corresponding to $(\Wen, \Ohd, \Ohf) = (4, 0.034, 0.67)$. Videos~\red{SM1--SM3} are available in the supplementary material at \red{X}.}
		\label{fig:figure02}
	\end{figure}
	
	In figure~\ref{fig:figure02}, we compare the behavior of a typical silicone oil drop ($R = 1.0\,\si{\milli\meter}$, $V = 0.35\,\si{\meter}/\si{\second}$, and $\eta_d = 4.6\,\si{\milli\pascal}.\si{\second}$, \revRev{\emph{i.e.} $(\Wen, \Ohd, \Bon) = (4, 0.034, 0.5)$}) impacting on films with a fixed viscosity $\eta_f = 96\,\si{\milli\pascal}.\si{\second}$ \revRev{($\Ohf = 0.67$)} but contrasting thicknesses: $h_{f} = 0.01,\,0.35$, and $0.85\,\si{\milli\meter}$ \revRev{(\emph{i.e.} $\Gamma = 0.01,\,0.35,$ and $0.85$, respectively)}. We show a one-to-one comparison between experimental and DNS snapshots, and display three key pieces of information: the position of the liquid-air interfaces (green lines) that can be directly compared with experiments, the rate of viscous dissipation per unit volume (left panel of each numerical snapshot), and the magnitude of the velocity field (right panel of each numerical snapshot).
	
	For the thinnest film ($h_f = 0.01\,\si{\milli\meter}$, figure~\ref{fig:figure02}a and supplementary video~\red{SM1}), the drop deforms as it comes in apparent contact with the film mediated by the air layer, an instant that we choose as the origin of time $t = 0$. The drop spreads until it reaches its maximal lateral extent, recoils and rebounds in an elongated shape after a time $t_c = 15.6 \pm 0.1\,\si{\milli\second}$, called the contact time. 
	Throughout the impact process, viscous stresses inside the drop dissipate energy (see times $t = 1.5\,\si{\milli\second}\,\text{and}\,7.5\,\si{\milli\second}$). Consequently, after take off, the drop reaches a maximal center of mass height $H = 2.0 \pm 0.1\,\si{\milli\meter}$ relative to the undisturbed film surface, from which we deduce the restitution coefficient defined as $\varepsilon = \sqrt{2g(H-R)}/V$, here $\varepsilon = 0.48\pm 0.05$. 
	The liquid--air interface profiles obtained from experiments and numerics are in excellent agreement, and we measure the same value of the contact time and restitution coefficient in simulations, using the method described in appendix~\ref{sec:restitution in simulations}. 
	This behavior is in quantitative agreement with that reported for the impact of a viscous drop on a superhydrophobic surface by \citet{jha2020viscous}, suggesting that the presence of both the air and liquid film have a negligible influence on the macroscopic dynamics of the rebound, and that viscous dissipation in the drop determines the rebound height.
	
	For $h_f = 0.35\,\si{\milli\meter}$ (figure \ref{fig:figure02}b and supplementary video~\red{SM2}), despite the noticeable deformation of the liquid film, the qualitative features of the bounce are similar. We further observe that as the drop takes off, the film free surface has not yet recovered its undisturbed position. We measure an increase of the contact time to $t_c = 17 \pm 0.1\,\si{\milli\second}$ and a decrease in the rebound elasticity, with $H = 1.6 \pm 0.1\,\si{\milli\meter}$ implying $\varepsilon = 0.37 \pm 0.04$.  The DNS snapshots show that in this case, viscous dissipation occurs both in the drop and the underlying liquid. Qualitatively, the instantaneous rate of viscous dissipation in the drop is similar for $h_f = 0.01\,\si{\milli\meter}$ and $h_f = 0.35\,\si{\milli\meter}$, suggesting that the decrease in rebound elasticity is primarily linked to the increased film dissipation.
	
	Finally, for $h_f = 0.85\,\si{\milli\meter}$ (figure \ref{fig:figure02}c and supplementary video~\red{SM3}), the film deformation increases and the substrate \revRev{loses} its repellent ability. The drop center of mass does not take off above $H=R$, the drop floats on top of the liquid film, a situation that corresponds to the inhibition of bouncing for which $\varepsilon \approx 0$ and the contact time diverges.
	In this case, we notice that the experimental and numerical interface profiles differ at $t= 0\,\si{\milli\second}$. This initial discrepancy, caused by drop oscillations upon detachment from the needle, does not affect the subsequent impact dynamics and the impact outcome, as evidenced by the good agreement of the interface profiles at later instants.
	
	\begin{figure}
		\centering
		\includegraphics[width=\textwidth]{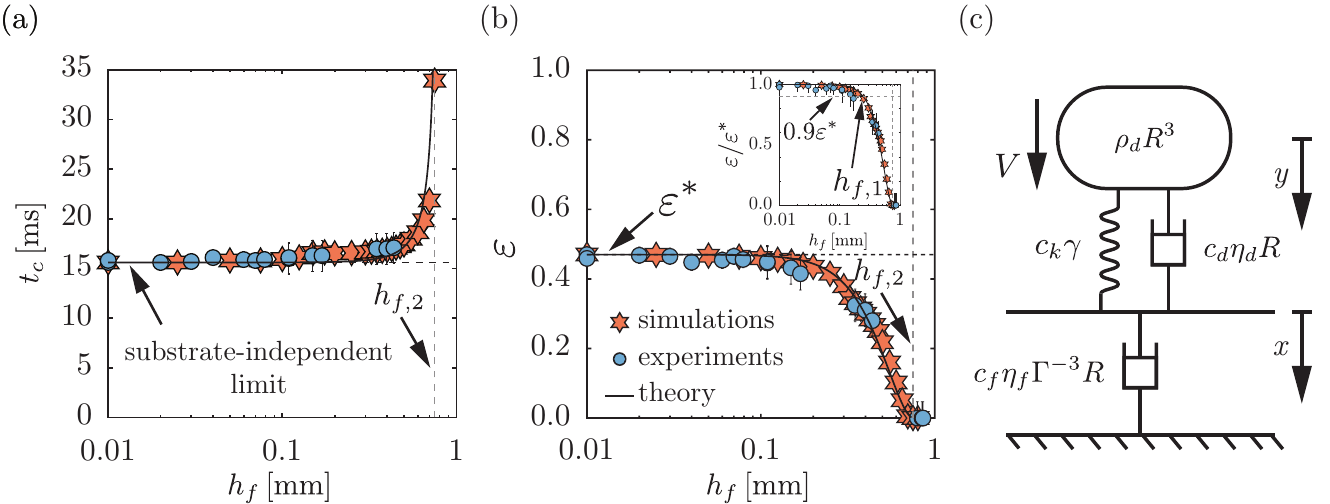}
		\caption{Effect of the film thickness on the rebound characteristics for $R = 1\,\si{\milli\meter}$, $V = 0.3\,\si{\meter}/\si{\second}$, $\eta_{d} = 4.6\,\si{\milli\pascal}.\si{\second}$ and $\eta_{f} = 96\,\si{\milli\pascal}.\si{\second}$, \emph{i.e.} $(\Wen, \Ohd, \Ohf) = (4, 0.034, 0.67)$: (a) contact time $t_{c}$ and (b) restitution coefficient $\varepsilon$ as a function of film thickness $h_{f}$. Circles and hexagrams represent experiments and DNS, respectively. 
			In panels (a) and (b), the horizontal black dashed lines represent the substrate-independent limits of contact time and restitution coefficient, respectively, while the solid black lines shows the results from the phenomenological model (see \S~\ref{sec:Phenomenological_model}) with parameters $c_{k} = 2$, $c_{d} = 5.6$ and $c_{f} = 0.46$. The vertical gray dotted line marks the transition from the bouncing to the non-bouncing (floating) regime. The inset of panel (b) illustrates the variation of the restitution coefficient normalized by its substrate-independent value $\varepsilon^*$ as a function of the film thickness. Here, the horizontal gray line represents $\varepsilon = 0.9\varepsilon^*$, marking the transition from substrate--independent to substrate--dependent bouncing at $h_f = h_{f, 1}$. (c) Schematic diagram of the phenomenological model that describes the drop impact process on a liquid film. The parameters $\rho_d R^{3}$, $\eta_{d}R$, and $\gamma$ are associated to the drop properties and $\eta_{f} \Gamma^{-3} R$ is associated to the film properties. }
		\label{fig:figure03}
	\end{figure}
	
	We now systematically vary the film thickness $h_{f}$ while keeping the drop and film viscosities constant ($\eta_d = 4.6\,\si{\milli\pascal}.\si{\second}$ and $\eta_f = 96\,\si{\milli\pascal}.\si{\second}$) and plot, in figures~\ref{fig:figure03}(a) and~(b), the contact time $t_c$ and the coefficient of restitution $\varepsilon$ extracted from experiments (circles) and  DNS (hexagrams).
	Experiments and simulations are in excellent agreement when varying the film thickness by two orders of magnitude, $h_{f} = 0.01\,\si{\milli\meter} - 1\,\si{\milli\meter}$. 
	The existence of two regimes is readily apparent. Firstly, for $h_{f} \lesssim 0.1\,\si{\milli\meter}$, both $t_{c}$ and $\varepsilon$ are independent of $h_f$. The value of the contact time in this regime, $t_c = 15.6 \pm 0.5\,\si{\milli\second}$, corresponds to that expected from the inertio-capillary scaling \citep{wachters1966heat, richard2002contact}. The contact time is proportional to $\tau_\gamma = \sqrt{\rho_d R^3/ \gamma}$ with a prefactor $2.2 \pm 0.1$, in good agreement with that calculated by \citet{rayleigh1879capillary} for the fundamental mode of drop oscillation $\pi/\sqrt 2$. Similarly, the plateau value of the coefficient of restitution $\varepsilon = 0.47 \pm 0.04$ is in reasonable agreement with that reported for the impact of water drops on superhydrophobic substrates for similar drop Ohnesorge number $\Ohd$ and impact Weber number $\Wen$ \citep{jha2020viscous}. We therefore refer to this regime as substrate--independent rebound (see also appendix~\ref{app:SubstrateIndependentBouncing}). 
	
	Secondly, for $h_{f} \gtrsim 0.1\,\si{\milli\meter}$, the contact time and coefficient of restitution are influenced by the film thickness. We observe that $t_c$ increases (figure \ref{fig:figure03}a) and $\varepsilon$ decreases (figure \ref{fig:figure03}b) with increasing $h_f$ until $t_c$ diverges and bouncing ceases ($\varepsilon = 0$) for $h_f \approx 0.75$ mm. This critical thickness marks the threshold of the rebound behavior and the transition to the non-bouncing (floating) regime. Here, the rebound characteristics vary significantly with $h_f$ and we therefore refer to this regime as substrate--dependent.
	
	Finally, we characterize the transition from the substrate--independent to the substrate--dependent regime by introducing the thickness $h_{f,1}$, in dimensionless form $\Gamma_1 = h_{f,1}/R$, which marks the decrease of $\varepsilon$ to $0.9$ times its plateau value $\varepsilon^*$. 
	Similarly, we define the critical thickness $h_{f,2}$, respectively $\Gamma_2 = h_{f,2}/R$, associated to the transition from the substrate--dependent to the non-bouncing (floating) regime as the smallest film thickness which results in $\varepsilon = 0$. 
	The impact dynamics can be categorized into three distinct regimes: a substrate--independent regime for $\Gamma = h_{f}/R \le \Gamma_1$, a substrate--dependent regime for $\Gamma_1 < \Gamma < \Gamma_2$, and a non-bouncing (floating) regime for $\Gamma \ge \Gamma_2$.
	
	\section{Phenomenological model} \label{sec:Phenomenological_model}
	
	We now seek to rationalize the dependence of the rebound time and elasticity with the substrate and drop properties by constructing a minimal model, guided by our experimental and numerical observations. We build on the classical description of a drop as a liquid spring which reflects the balance of inertia and capillarity during a rebound \citep{richard2002contact,okumura2003water}. Here, we consider viscous drops and further add a damping term to the liquid spring, an approach which has been shown to successfully capture the variation of contact time and coefficient of restitution across over two orders of magnitude variation in liquid viscosities \citep{jha2020viscous}. Similarly, we interpret the film behavior through the liquid spring analogy. The film motion contrasts with that of the drop, while the latter displays a full cycle of oscillation during a rebound, the former never returns to its undisturbed position (see figure~\ref{fig:figure02} and supplementary movies \red{SM1} - \red{SM3}). This observation leads us to consider that the damping component dominates the behavior of the liquid film, and to neglect the contributions of inertia and surface tension. We further discuss this assumption and its validity in \S~\ref{sec:EnergySection}. 
	
	In figure \ref{fig:figure03}(c), we present a sketch of the model, where we assume that the droplet and the film are connected in series during apparent contact, and show the scaling forms of the drop and film components. The scaling relations for the drop mass, stiffness and damping are taken from the work of \citet{jha2020viscous} as proportional to \revRev{$\rho_d R^3$, $\gamma$}, and $\eta_d R$, respectively, with corresponding prefactors of $1$, $c_k$, and $c_d$. We determine the values of $c_k$ and $c_d$ from results in the substrate--independent bouncing regime (see appendix~\ref{app:SubstrateIndependentBouncing}).
	The scaling form of the film damping term is chosen as proportional to $\eta_{f} \Gamma^{-3} R$, where $\Gamma = h_f/R$, with corresponding prefactor of $c_{f}$ (figure~\ref{fig:figure03}c). 
	This is built on two key assumptions. First, we assume that the lubrication approximation holds in the film as, for sufficiently high film Ohnesorge numbers ($\Ohf \gtrsim 0.1$), the slopes associated to the film deformations are small ($\Gamma \ll 1$, $\Ohf \sim \mathcal{O}\left(1\right)$, see \S~\ref{sec:EnergySection} for limitations).
	And second, we choose to consider the drop as an impacting disk rather than a sphere, owing to the rapid drop spreading upon impact \citep{eggers2010drop, wildeman-2016-jfm}, which results in a damping term proportional to $\Gamma^{-3}$ instead of $\Gamma^{-1}$ \citep{leal2007advanced}. Lastly, we fit the prefactor $c_f$ to our experiments and simulations.
	
	With these assumptions, the equations of motion for the model system (figure~\ref{fig:figure03}c) read
	
	\begin{align}
		\revRev{\rho_d} R^{3} \ddot{y} &= -c_{k} \gamma \left( y - x \right) - c_{d} \eta_{d} R \left( \dot{y} - \dot{x} \right), \label{sec5_eq01a} \\
		0 &= +c_{k} \gamma \left( y - x \right) + c_{d} \eta_{d} R \left( \dot{y} - \dot{x} \right) - c_{f} \eta_{f} \Gamma^{-3} R \dot{x}, \label{sec5_eq01b}
	\end{align}
	
	\noindent where $y$ and $x$ are the displacements of the drop and the film relative to their initial position in the reference frame of the laboratory, and the dots denote time derivatives. 
	We point out that by setting $\dot{x} = x = 0$, we recover the model proposed by \citet{jha2020viscous}, which extends the analogy between the drop impact process and a spring-mass system \citep{okumura2003water} by adding a damper to account for viscous dissipation in the drop. Here, we additionally consider viscous dissipation in the liquid coating and model the film as a damper without inertia. 
	\revRev{We make this modelling assumption, guided by the overdamped dynamics of the film (figure \ref{fig:figure02}), to keep the number of free parameters to as few as possible (three in the present study, $c_k$, $c_d$, and $c_f$).
		We stress here that $c_k$ and $c_d$ are fixed in this study, and that their value is in quantitative agreement with the corresponding prefactors derived by \citet{jha2020viscous}.}
	
	Similarly as for the governing equations in DNS, we make equations~\eqref{sec5_eq01a} and~\eqref{sec5_eq01b} dimensionless using the length scale $R$ and the time scale $\tau_{\gamma}$ and use tildes to identify dimensionless variables. Next, we obtain an equation of motion for the drop deformation $\tilde{z} = \tilde{y} - \tilde{x}$, namely
	
	\begin{align}
		\left( 1 + \frac{c_{d} \Ohd}{c_{f} \Ohf \Gamma^{-3}} \right) \ddot{\tilde{z}} + c_{d} \Ohd \left( 1 + \frac{c_{k}}{c_{d} \Ohd \cdot c_{f} \Ohf \Gamma^{-3}} \right) \dot{\tilde{z}} + c_{k} \tilde{z} = 0, \label{sec5_eq02}
	\end{align}
	
	\noindent which admits oscillatory solutions, that is drop rebound, under the condition
	
	\begin{align}
		\omega^2 = 4 c_{k} - \left( c_{d} \Ohd  - \dfrac{c_{k}}{c_{f} \Ohf \Gamma^{-3}} \right)^{2} > 0.
		\label{eqn:sec1_Omega2}
	\end{align}
	
	\noindent We note that $\omega^2$ decreases with increasing $\Gamma$ for fixed $\Ohd$ and $\Ohf$, in qualitative agreement with the existence of a critical film height above which bouncing stops (figure \ref{fig:figure03}b).
	Equation \eqref{eqn:sec1_Omega2} allows us to determine the bounds of the bouncing regime in terms of a critical drop Ohnesorge number $Oh_{d,c}$ and film thickness $\Gamma_2$. Discarding the two roots of the equation $\omega^2 = 0$ that yield unphysical negative values of $Oh_{d,c}$ and $\Gamma_2$, we obtain
	
	\begin{align}
		\label{eqn:omega0_2}
		Oh_{d,c} &= \frac{1}{c_d}\left(2\sqrt{c_k} + \frac{c_{k}}{c_{f}}\left(\Gamma_2/\Ohf^{1/3}\right)^3\right),\,\text{and}\\
		\label{eqn:omega0_1}
		\Gamma_2/\Ohf^{1/3} &= \left(\frac{c_{f}}{c_{k}}\left( c_{d} \Ohd + 2 \sqrt{c_{k}} \right)\right)^{1/3}.
	\end{align}
	
	\noindent Equations~\eqref{eqn:sec1_Omega2}--\eqref{eqn:omega0_1} evidence that the role of the film viscosity and height are intertwined as we find the combination $\Gamma/\Ohf^{1/3}$ \revRev{that can be inferred as the effective film thickness or mobility}.
	Furthermore, the substrate--independent bouncing threshold is recovered when \revRev{this} film mobility, $\Gamma/\Ohf^{1/3}$, tends to 0, that is for very thin and/or very viscous films. 
	Indeed, equations~\eqref{eqn:omega0_2}--\eqref{eqn:omega0_1} become
	
	\begin{align}
		\label{eqn:omega1_2}
		Oh_{d,c} &= \frac{2\sqrt{c_k} }{c_d},\,\text{and}\\
		\label{eqn:omega1_1}
		\Gamma_2/\Ohf^{1/3} &= \left(2\frac{c_{f}}{\sqrt{c_{k}}}\right)^{1/3},
	\end{align}
	
	\noindent for the limiting cases of substrate--independent ($\Gamma/\Ohf^{1/3} \to 0$), and inviscid drop ($\Ohd \to 0$) asymptotes, respectively.
	
	To go further, we solve equation~\eqref{sec5_eq02} with the initial conditions $\tilde{z} = 0$ and $\dot{\tilde{z}} = \sqrt{\Wen}$ at $\tilde{t} = 0$, yielding
	
	\begin{align}
		\label{eqn:displacement_Theory}
		&\tilde{z}(\tilde{t}) =  \frac{2 \sqrt{\Wen}} {\Omega}\exp\left(-\frac{\mathcal{\phi} \tilde{t}}{2}\right) \sin\left(\frac{\Omega\tilde{t}}{2}\right),\\
		\label{eqn:displacement_damper}
		&\text{where}\quad\mathcal{\phi} = \frac{c_k + c_d\Ohd c_f\Ohf\Gamma^{-3}}{c_d\Ohd+c_f\Ohf\Gamma^{-3}}\\
		\label{eqn:displacement_pulsation}
		&\text{and}\quad\Omega = \omega\left(1+\frac{c_d\Ohd}{c_f\Ohf\Gamma^{-3}}\right)^{-1}
	\end{align}
	
	\noindent can be interpreted as an effective damper and angular frequency, respectively, by comparing the above expression to the one obtained by \citet{jha2020viscous} for $\Gamma/\Ohf^{1/3} \to 0$. We can deduce the expressions for both the contact time and the coefficient of restitution using these pieces of information. The contact time is taken as the instant at which the drop deformation $\tilde{z}$ comes back to zero, which occurs at $\Omega \tilde{t} = 2\pi$, giving
	
	\begin{align}
		\frac{t_{c}}{\tau_{\gamma}} =  \dfrac{2 \pi}{\omega} \left( \dfrac{c_{d} \Ohd}{c_{f} \Ohf\Gamma^{-3}} + 1 \right). \label{eqn:tc_model}
	\end{align}
	
	\noindent Equation~\eqref{eqn:tc_model} is then used to compute the coefficient of restitution $\varepsilon$ as the ratio of rebound velocity, $\dot{\tilde{z}}(\tilde{t}_c)$, to the impact velocity, $\sqrt{\Wen}$.
	We immediately notice that this definition yields an expression for $\varepsilon$ that does not depend on $\Wen$, in contrast with the experimentally observed decrease of $\varepsilon$ with $\Wen$.
	\revRev{We account for the Weber number dependence of $\varepsilon$, which is not captured by spring mass models \citep{jha2020viscous}, by scaling the coefficient of restitution by $\varepsilon_0(\Wen)$, its $\Wen$--dependent value in the substrate--independent limit for inviscid drops,}
	
	\begin{align}
		\varepsilon(\Wen, Oh_d, Oh_f, \Gamma)  = \varepsilon_0(\Wen) \exp\left( -\dfrac{\pi}{\omega} \left( c_{d} \Ohd + \dfrac{c_{k}}{c_{f} \Ohf \Gamma^{-3}} \right) \right),
		\label{eqn:eps_model}
	\end{align}
	
	\noindent \revRev{where the prefactor $\varepsilon_0(\Wen)$ is not a model prediction. We obtain the other prefactors, $c_k\,\&\,c_d$ by fitting the substrate--independent experiments following \citet{jha2020viscous}. 
		This simplification allows us to recover the expressions for $t_c$ and $\varepsilon$ for viscous drop impact on non-wetting substrates \citep{jha2020viscous}, and thus to determine $c_k = 2$ and $c_d = 5.6$ which we keep fixed during this study. For details of this simplification and on the determination of the prefactors, see appendix~\ref{app:SubstrateIndependentBouncing}.}
	
	We test the model predictions for the contact time and rebound elasticity in the substrate--dependent regime by comparing the data (symbols) presented in figures \ref{fig:figure03}(a) and (b) to least-square fits of equations~\eqref{eqn:tc_model} and~\eqref{eqn:eps_model} with $c_f$ as a free parameter (solid lines) and taking $\varepsilon_0 = 0.58$ (see appendix~\ref{app:SubstrateIndependentBouncing}). We find that the model accurately predicts the variation of $t_c$ and $\varepsilon$ with $\Gamma$ for $c_f = 0.46 \pm 0.1$. \revRev{For the rest of this work, we fix $c_f = 0.46$ and assess the predictive ability of the simplified model.}
	
	
	\section{Influence of drop and film parameters}\label{sec:influence}
	
	\revRev{We now test the model predictions and limits by experimentally and numerically varying the drop and film Ohnesorge numbers. We give particular attention to the value of the coefficient $c_f$ (fixed at $0.46$) necessary to fit the model to this data, and to the two asymptotes predicted by the model that bound the bouncing domain (equations~\eqref{eqn:omega1_2}-\eqref{eqn:omega1_1}).}
	
	\subsection{Influence of the film Ohnesorge number $\Ohf$}
	
	\begin{figure}
		\centering
		\includegraphics[width=\linewidth]{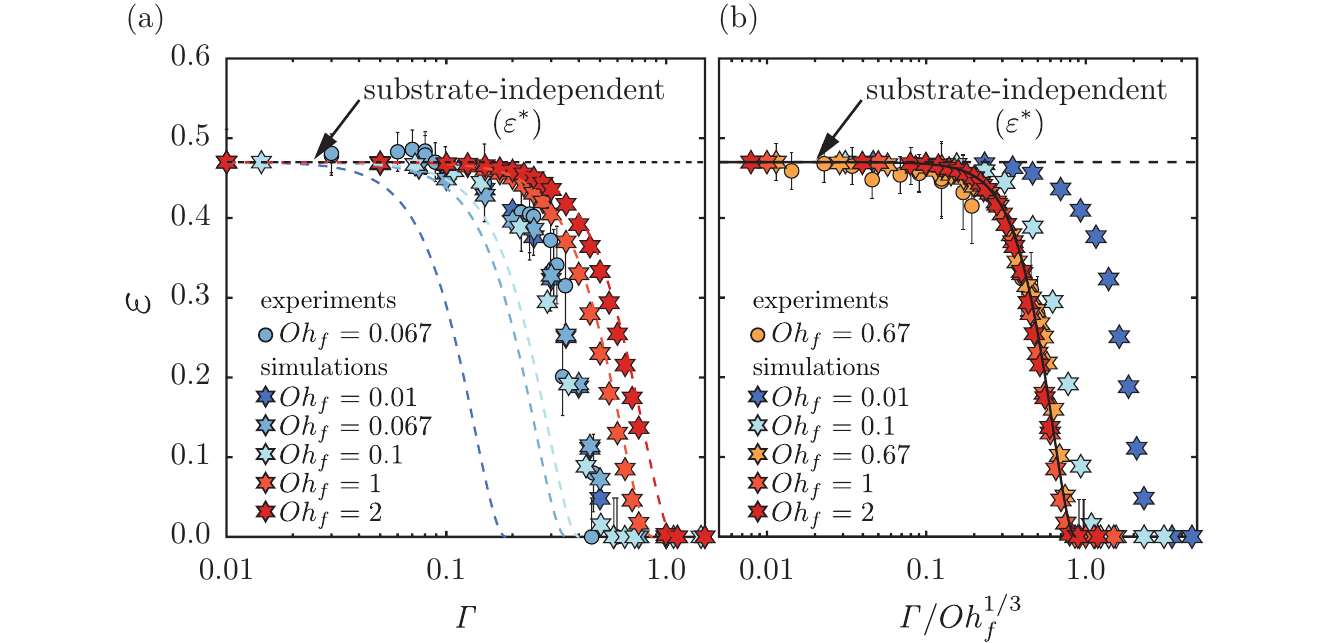}
		\caption{Influence of the film parameters on the impact characteristics: variation of the coefficient of restitution $\varepsilon$ as a function of (a) the film thickness $\Gamma$ and (b) the film mobility $\Gamma / \Ohf^{1/3}$. In panels (a) and (b), the circles and hexagrams correspond to the results from experiments and simulations, respectively. The colored dashed lines in panel (a) and the solid black line in panel (b) illustrate the results from the phenomenological model (equation~\eqref{eqn:eps_model}) with parameters $c_{k} = 2$, $c_{d} = 5.6$ and $c_{f} = 0.46$. Black dashed lines in panels (a) and (b) mark the substrate--independent limit of the restitution coefficient $\varepsilon^*$. For all cases in this figure, $\Ohd = 0.034$ and $\Wen = 4$.}
		\label{fig:controlParameters1}
	\end{figure}
	
	We first vary the film Ohnesorge number $\Ohf$ while keeping the drop and impact properties constant. In figure \ref{fig:controlParameters1}(a), we show the evolution of the coefficient of restitution $\varepsilon$ for drops with $\Ohd = 0.034$ as a function of the dimensionless film thickness $\Gamma$ while exploring two decades in film viscosity, $\Ohf = 0.01 - 2.0$. On the one hand, as expected, the values of the coefficient of restitution are not affected in the substrate--independent limit. On the other hand, the substrate--dependent behavior shows the influence of $\Ohf$ and we identify two regimes. For $\Ohf < 0.1$, the evolution of $\varepsilon$ with $\Gamma$ does not depend on $\Ohf$, as illustrated by the data collapse in figure \ref{fig:controlParameters1}(a). However, for $\Ohf > 0.1$, increasing the film viscosity leads to a larger extent of the substrate--independent plateau and to an increase of the critical film thickness at which bouncing stops. This change in the $\Ohf$ dependence can be characterized by the two dimensionless critical film thicknesses $\Gamma_1 = h_{f,1}/R$, and $\Gamma_2 = h_{f,2}/R$, which increase from $0.17$ to $0.33$ and $0.58$ to $1.1$, respectively, when $\Ohf$ is increased from $0.1$ to $2.0$. 
	
	We interpret the two types of behavior in the substrate--dependent regime in the light of our minimal model, which predicts that the film mobility, $\Gamma/\Ohf^{1/3}$, controls the dissipation in the substrate. 
	In figure \ref{fig:controlParameters1}(b), we plot the coefficient of restitution data presented in (a) after rescaling the horizontal axis by $\Ohf^{-1/3}$. 
	The data now collapse for \revRev{$\Ohf > 0.1$}, indicating that the proposed approximations capture the large viscosity limit but breaks down for lower film Ohnesorge numbers. We further evidence the validity and failure of the minimal model by plotting the predictions of equation~\eqref{eqn:eps_model} with $c_f = 0.46$ (\revRev{dashed colored lines in figure \ref{fig:controlParameters1}a and solid black line in figure \ref{fig:controlParameters1}b}). \revRev{The minimal model predicts the restitution coefficient accurately for $Oh_f > 0.1$, suggesting that our modelling assumptions are valid in this regime: the liquid film dynamics is dominated by viscous dissipation, and the flow can be successfully modelled in the lubrication approximation by assimilating the impacting drop to a cylinder.}

\subsection{Influence of the drop Ohnesorge number $\Ohd$}
\begin{figure}
	\centering
	\includegraphics[width=\textwidth]{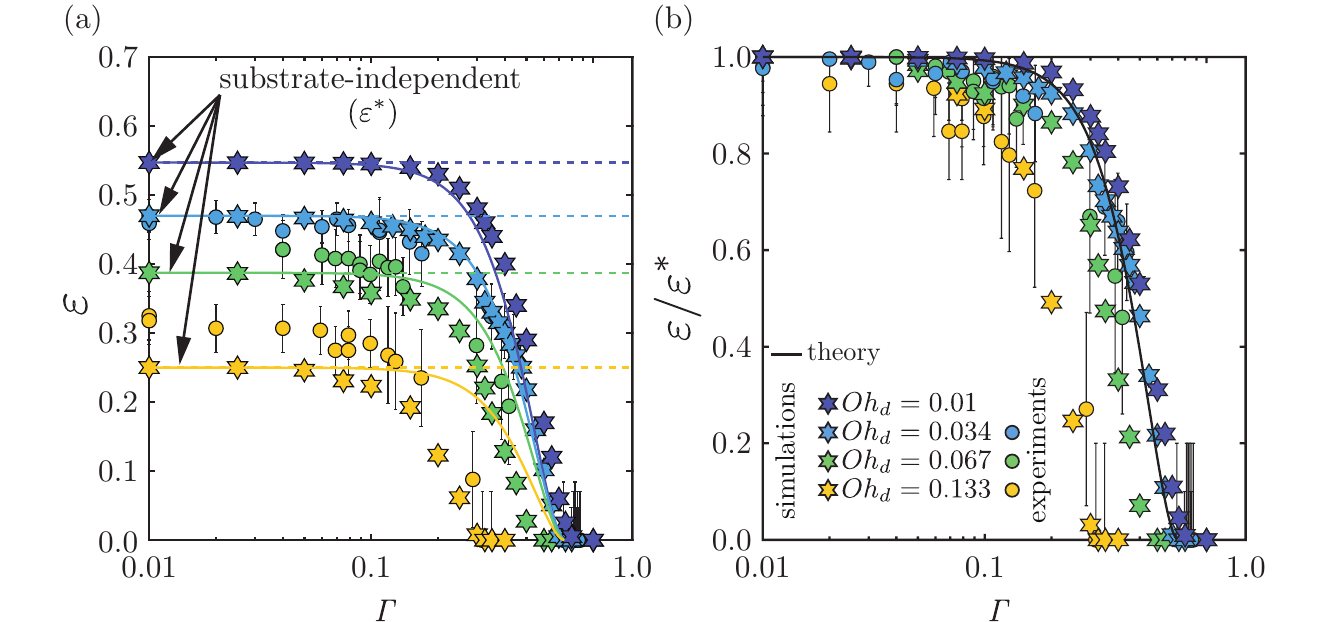}
	\caption{Influence of the drop parameters on the rebound elasticity: variation of (a) the coefficient of restitution $\varepsilon$ and (b) the coefficient of restitution normalized with its substrate--independent value $\varepsilon/\varepsilon^*$ as a function of the normalized film thickness $\Gamma$. The circles and hexagrams correspond to the results from the experiments and simulations, respectively. In panel (a), the dashed lines denote the plateau values of the restitution coefficient $\varepsilon^*(\Ohd)$. In panels (a) and (b), the solid lines represent the results from the phenomenological model (equation~\eqref{eqn:eps_model}) with parameters $c_{k} = 2$, $c_{d} = 5.6$ and $c_{f} = 0.46$. For all cases in this figure, $\Ohf = 0.667$ and $\Wen = 4$.}
	\label{fig:controlParameters2}
\end{figure}

In this section, we focus on the influence of the drop Ohnesorge number on the rebound elasticity.
In figure \ref{fig:controlParameters2}(a), we plot the coefficient of restitution as a function of the dimensionless film thickness for a fixed $\Ohf = 0.667$ and for varying $\Ohd$ spanning the range $0.01 - 0.133$. Increasing $\Ohd$ affects $\varepsilon$ across all film thicknesses. In the substrate--independent region, the coefficient of restitution decreases with increasing drop Ohnesorge number. In appendix~\ref{app:SubstrateIndependentBouncing}, we show that the plateau values reported in figure \ref{fig:controlParameters2}(a) decay exponentially with increasing $\Ohd$ as predicted by \citet{jha2020viscous}. To better illustrate the influence of $\Ohd$ in the substrate--dependent regime, we normalize the coefficient of restitution $\varepsilon$ by its substrate--independent value $\varepsilon^*$ (figure \ref{fig:controlParameters2}b). 
\revRev{With this normalization, we expect the data to follow the prediction of equation \eqref{eqn:eps_model} (solid line). 
	The data collapse only for small $\Gamma$, indicating that the phenomenological model predicts the influence of $\Ohd$ \revRev{only} in the substrate-independent limit. 
	This suggests that the model fails to account for the interplay between the drop and the film properties that affects dissipation in both liquids, and ultimately the coefficient of restitution.
	Here as well, we monitor the $\Ohd$ dependence of the coefficient of restitution, and its deviation from the prediction of equation \eqref{eqn:eps_model},} through the evolution of $\Gamma_1$ and $\Gamma_2$, that both decrease with increasing Ohnesorge number.

\subsection{Influence of $\Ohf$ and $\Ohd$ on the critical film thicknesses}
\begin{figure}
	\centering
	\includegraphics[width=\textwidth]{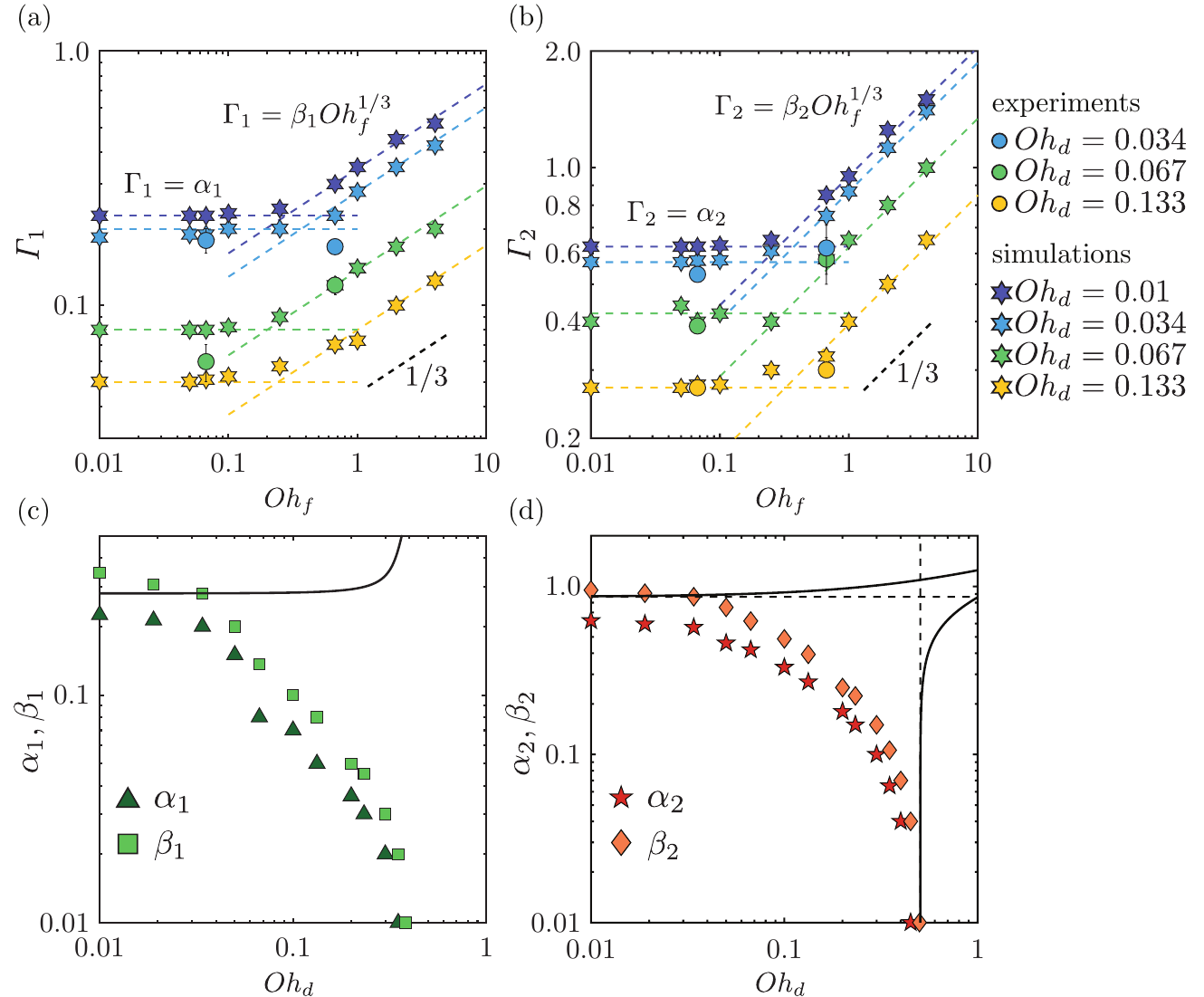}
	\caption{Critical film thickness marking the transition from (a) substrate--independent to substrate--dependent bouncing $\Gamma_1$ and (b) bouncing to non-bouncing (floating) $\Gamma_2$ as a function of $\Ohf$ at different $\Ohd$. 
		Prefactors (c) $\alpha_{1}$ and $\beta_{1}$, and (d) $\alpha_{2}$ and $\beta_{2}$ as a function of $\Ohd$. 
		The solid black line in panel (c) represents the model prediction for $\beta_1$, equation~\eqref{alpha1}. 
		The solid black lines in panel (d) represent the model predictions for $\beta_2$ using equations~\eqref{eqn:omega0_2}-\eqref{eqn:omega0_1}, and the black dashed lines show the two asymptotes, equations~\eqref{eqn:omega1_2}-\eqref{eqn:omega1_1}.}
	\label{fig:Gamma1Gamma2}
\end{figure}

We now \revRev{quantify} the influence of the drop and film Ohnesorge number by reporting their effect on the critical thicknesses for substrate--independent to substrate--dependent ($\Gamma_1$) and bouncing to non-bouncing (floating) ($\Gamma_2$) transitions. Indeed, we have shown above that these two critical thicknesses are good proxies to characterize the continuous transition from substrate--independent bouncing to rebound inhibition. In figures~\ref{fig:Gamma1Gamma2}(a,b), we show $\Gamma_1$ and $\Gamma_2$ as a function of the film Ohnesorge number for $\Ohd$ in the range $0.01 - 0.133$. This representation reflects the existence of the two distinct regimes reported in figure~\ref{fig:controlParameters1}. 

First, when $\Ohf < 0.1$, $\Gamma_1$ and $\Gamma_2$ are independent of $\Ohf$. We write $\Gamma_1 = \alpha_1(\Ohd)$ and $\Gamma_2 = \alpha_2(\Ohd)$, \revRev{and report the values of $\alpha_1(\Ohd)$ and $\alpha_2(\Ohd)$ in figures \ref{fig:Gamma1Gamma2}(c,d)}. 
This observation is in contradiction with the expectations from our minimal model which predicts that $\Ohf$ influences the values of $\Gamma_1$ and $\Gamma_2$. Surprisingly, this $\Ohf$--independence of the critical thicknesses and the collapse observed in figure \ref{fig:controlParameters1}(a) suggest that the energy transfer to the film (in the form of kinetic and surface energies) and the film viscous dissipation are independent of film viscosity for $\Ohf < 0.1$. We will further elaborate on this regime in \S~\ref{sec:EnergySection}. 

Second, for larger film Ohnesorge numbers, the dissipation in the film is captured by the lubrication approximation ansatz. As a result, both critical thicknesses follow the relations $\Gamma_1 = \beta_1(\Ohd)\Ohf^{1/3}$ and $\Gamma_2 = \beta_2(\Ohd)\Ohf^{1/3}$, as predicted by the model. 
Beyond this scaling relation, the accuracy of the minimal model is tied to its ability to predict the prefactors $\beta_1$ and $\beta_2$ when $\Ohf \gtrsim 0.1$. In figures~\ref{fig:Gamma1Gamma2}(c,d), we plot $\beta_1$ and $\beta_2$ as a function of the drop Ohnesorge number. Both prefactors show a plateau for $\Ohd \lesssim 0.03$ before decreasing monotonically with the drop Ohnesorge number.
We compare the measured prefactors to the model predictions which we plot as a solid lines in figures ~\ref{fig:Gamma1Gamma2}(c,d). 
$\beta_1$ is obtained by solving $\varepsilon = 0.9\varepsilon^*$, yielding

\begin{align}
	\beta_{1} = c_{f}^{1/3} \left[ \dfrac{-c_{d} \Ohd (1 - r^{2}) + 2 r \sqrt{c_{k} (1 + r^{2}) - c_d^{2} \Ohd^{2}}}{c_{k} (1 + r^{2})} \right]^{1/3}, \label{alpha1}\\
	\text{where, } r = \dfrac{c_{d} \Ohd}{\sqrt{4 c_{k} - c_{d}^{2} \Ohd^{2}}} - \dfrac{\text{ln}(0.9)}{\pi},
\end{align}

\noindent and $\beta_2$ is given by equation \eqref{eqn:omega0_1}.
The model fails to capture both the decrease of $\beta_1$ and $\beta_2$ with $\Ohd$. Yet, we can interpret the evolution of these two prefactors along the inviscid and viscous drop limiting cases.
Indeed, for inviscid drops (\emph{i.e.} small $\Ohd$), the model predictions for $\beta_1$ and $\beta_2$ show a plateau whose value is in good agreement with that reported in experiments. 
Conversely, for viscous drops (\emph{i.e.} large $\Ohd$), $\beta_2$ decreases with $\Ohd$ to match the asymptote associated to the substrate--independent bouncing inhibition occurring at $Oh_{d,c} \approx 0.5$ (equation \eqref{eqn:omega0_2} and dotted line in figure ~\ref{fig:Gamma1Gamma2}d).

We stress that the model predictions shown in figures~\ref{fig:Gamma1Gamma2}(c,d) consider a unique value of $c_f = 0.46 \pm 0.1$, determined from least-square fit in \S~\ref{sec:Phenomenological_model}. 
We attribute the failure of the model to predict the dependence on $\Ohd$ away from the two asymptotes to its simplified representation of the drop--film interactions. 
While these oscillator based models remarkably predict the global outcome of a rebound, that is, for example, the contact time, coefficient of restitution, and the bounds of bouncing, they fail at accurately representing the interactions, such as the drop or film deformations (equation~\eqref{eqn:displacement_Theory}), and their dynamics. 
For example, the force associated to drop impact is maximal at early times, when the drop shape is spherical, while the force exerted by a spring is proportional to deformation. 

\revRev{Our phenomenological model successfully captures the behavior of the liquid film for $\Ohf \ge 0.1$, giving support to the modeling assumptions. Within this regime, the model allows for a quantitative prediction of the contact time and coefficient of restitution for both inviscid (\emph{i.e.} low $\Ohd$) and viscous (\emph{i.e.} large $\Ohd$) drops, for a fixed set of constants $c_k$, $c_d$ and $c_f$. In between these two limits, the model fails to predict the $\Ohd$ dependence, a fact we attribute to the simplicity of the representation of drop-film interactions.}
More intriguingly, the minimal model also breaks down for $\Ohf \lesssim 0.1$, where we observe that the coefficient of restitution does not depend on the film Ohnesorge number. We will demystify this behavior in the next section. 

\section{Bouncing inhibition on low Ohnesorge number films}
\label{sec:EnergySection}

\begin{figure}
	\centering
	\includegraphics[width=\textwidth]{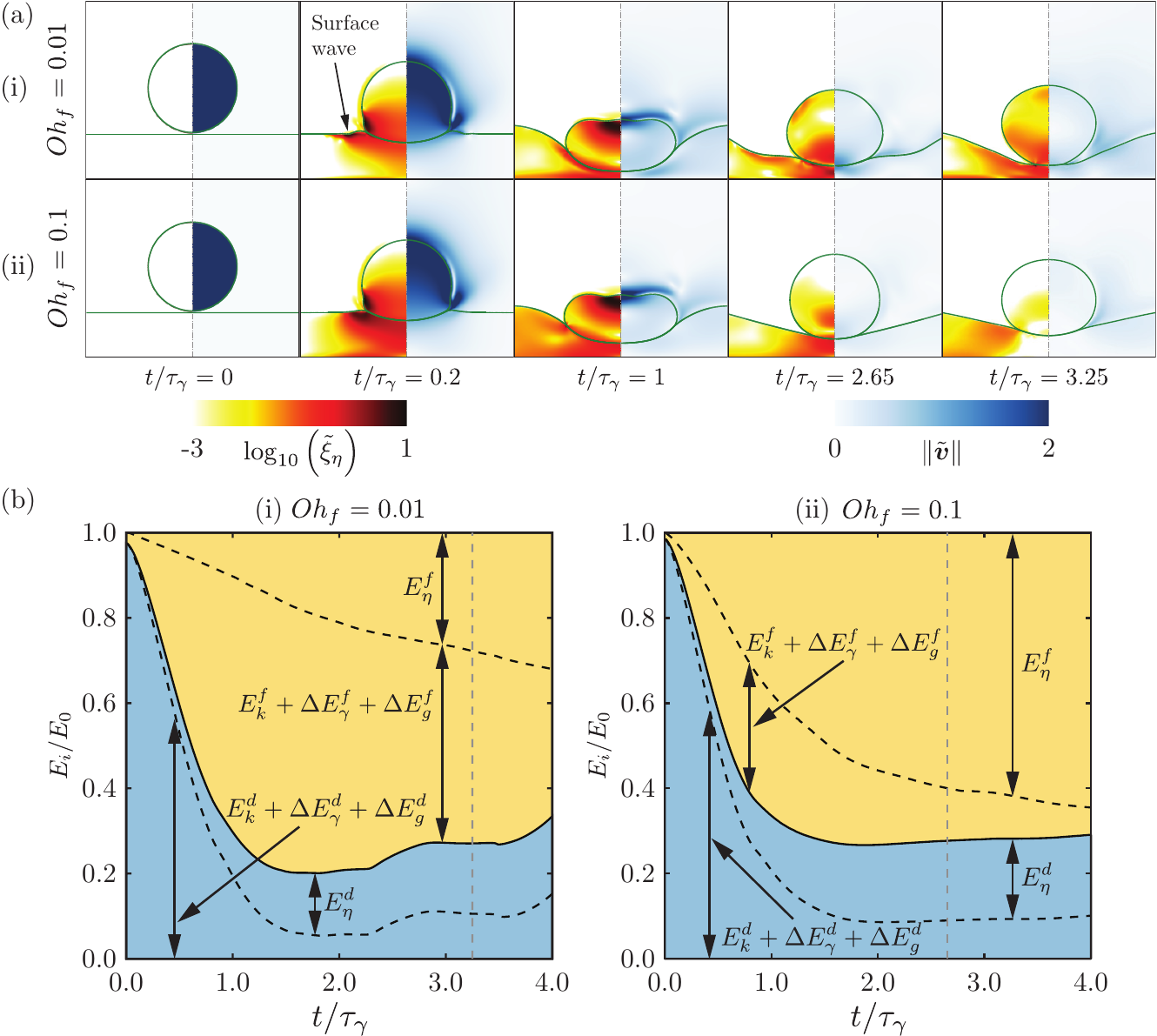}
	\caption{$\Ohf$ independent inhibition of bouncing: (a) typical drop impact dynamics on low viscosity films. The snapshots show the dimensionless rate of viscous dissipation per unit volume 
		on the left and the magnitude of dimensionless velocity field 
		on the right. 
		(b) Energy budgets for the two representative cases shown in panel (a), normalized by the available energy at the instant of impact. Here, the subscripts $g, k, \gamma,$ and $\eta$ denote gravitational potential, kinetic, surface, and viscous dissipation energies, respectively. The superscripts $d, f,$ and $a$ represent drop, film and air, respectively. The grey dashed line in each panel marks the instant when the normal reaction force between the drop and the film is minimum and represents the last time instant when the drop could have bounced off the film. In each panel, $\Ohf$ = (i) $0.01$ and (ii) $0.1$. For all the cases, $\Wen, \Ohd, \Gamma$ = $4, 0.034, 1$. }
	\label{fig:Plateau}
\end{figure}
\begin{figure}
	\centering
	\includegraphics[width=\textwidth]{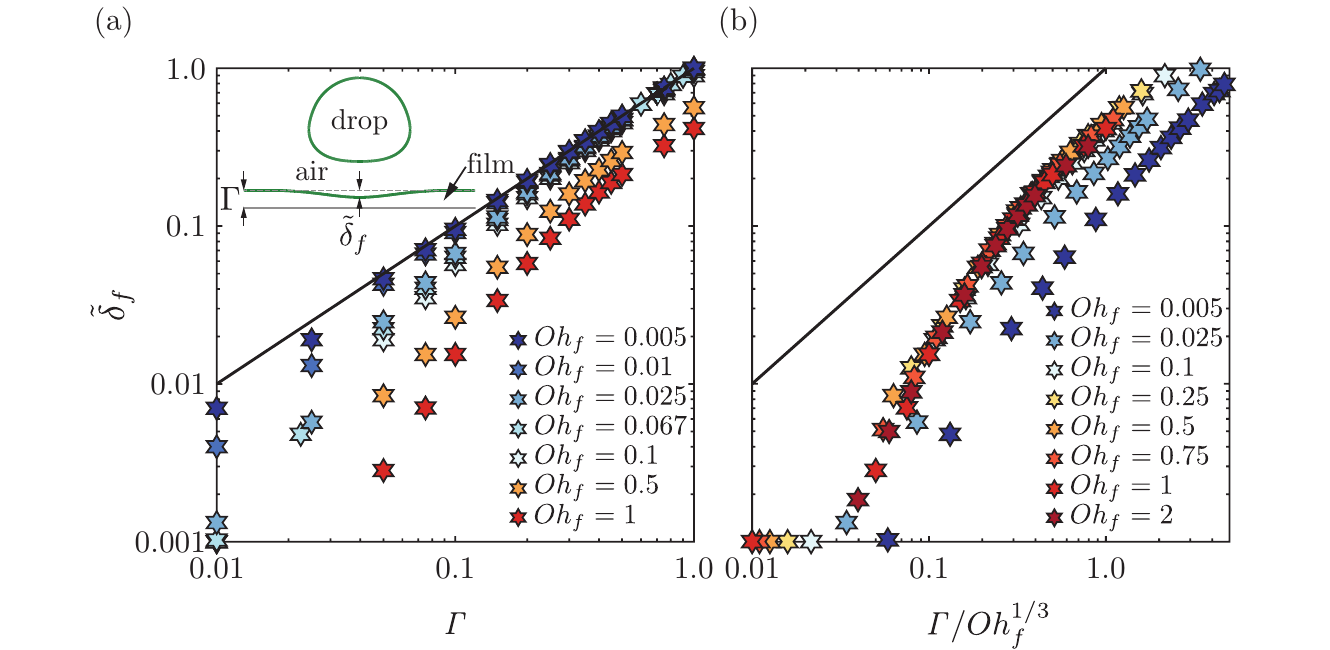}
	\caption{Dimensionless film deflection $\tilde{\delta}_f = \delta_f/R$ measured from the initial film free-surface (see the inset in panel a) as a function of (a) the film thickness $\Gamma$ and (b) the film mobility $\Gamma / \Ohf^{1/3}$ in the direct numerical simulations. 
		The solid black lines represent $\delta_f/R = \Gamma$ and $\delta_f/R = \Gamma/\Ohf^{1/3}$ in panels (a) and (b), respectively. For all cases in this figure, $\Ohd = 0.034$ and $\Wen = 4$.}
	\label{fig:Deflection}
\end{figure}

We now investigate the independence of the rebound elasticity with the film Ohnesorge number, illustrated by the data collapse of figure \ref{fig:controlParameters1}(a), for $\Ohf  < 0.1$.
Figure~\ref{fig:Plateau}(a) shows two typical impact scenarios in this regime, with $\Ohf = 0.01$ (figure~\ref{fig:Plateau}(a - i)) and $0.1$ (figure~\ref{fig:Plateau}(a - ii)), where bouncing is inhibited by the presence of the liquid film. 
Although these two representative cases differ by an order of magnitude in $\Ohf$, qualitatively, the drop shape and flow anatomy remain similar (figure~\ref{fig:Plateau}a, $t/\tau_\gamma = 0.2, 1$), suggesting an equal loading on the film. 
Nonetheless, the film response varies. 
We observe \revRev{surface waves} post impact on the film-air interface for $\Ohf = 0.01$, which \revRev{vanish} for $\Ohf = 0.1$ owing to increased \revRev{bulk} viscous attenuation (figure~\ref{fig:Plateau}a, $t/\tau_\gamma = 0.2, 2.65$). 

To further elucidate the drop--film interaction, we compute the energy budgets associated to the two representative cases with $\Ohf = 0.01$ (figure~\ref{fig:Plateau}(b - i)) and $0.1$ (figure~\ref{fig:Plateau}(b - ii)). 
The overall energy budget reads

\begin{align}
	E_0 =  \left(E_k^d + \Delta E_\gamma^d + \Delta E_g^d\right) + E_\eta^d + \left(E_k^f + \Delta E_\gamma^f + \Delta E_g^f\right)  + E_\eta^f + E_t^a,
	\label{eqn:EnergyBudgetGlobal}
\end{align} 

\noindent where $E_0$ is the energy at impact (i.e., the sum of the drop's kinetic and gravitational potential energies). The subscripts $g, k, \gamma,$ and $\eta$ denote gravitational potential, kinetic, surface, and viscous dissipation energies, respectively. Moreover, the superscripts $d, f,$ and $a$ represent drop, film and air, respectively. Lastly, reference values to calculate $\Delta E_g$ and $\Delta E_\gamma$ are at minimum $E_g$, and $E_\gamma$, at $t = 0$, respectively. Note that the contribution of the total energy associated with air ($E_t^a = E_k^a + E_\eta^a$) is negligible compared to other energies ($E_t^a(t/\tau_\gamma = 4) \approx 0.01E_0$). Readers are referred to \citet{landau2013course, wildeman-2016-jfm, ramirezsoto-2020-sciadv, sanjay2022taylor} for details of energy budget calculations. 

In both cases, the magnitude of the drop energy (the sum of the drop's kinetic, gravitational potential and surface energies) at the end of the rebound cycle, that is for $t = 3.25\tau_\gamma$ when $\Ohf = 0.01$ and $t = 2.65\tau_\gamma$ when $\Ohf = 0.1$ (vertical grey dashed lines), is similar, as expected from the independence of $\varepsilon$ with $\Ohf$. 
Note that the end of the cycle has been determined from the instant at which the reaction force between the drop and the film is minimum \citep[see appendix~\ref{sec:restitution in simulations} and][]{zhang2022impact}. Moreover, the energy budget evidences that the viscous dissipation in the drop during the rebound is similar, indicating that the magnitude of the energy transferred from the impacting drop to the film (the sum of the film's kinetic, gravitational potential and surface energies, and viscous dissipation) is not affected by the one order of magnitude change in $\Ohf$. 
Yet, the distribution of the film energy is dramatically different in the two cases we consider. For $\Ohf$ = 0.1, the energy transferred to the film is mostly lost to viscous dissipation, while for $\Ohf = 0.01$ the energy stored in the film's kinetic, surface and potential components dominates. 
We stress here that the $\Ohf$--independent behavior does not imply that dissipation is negligible. Indeed, the viscous dissipation in the film accounts for approximately $40\%$ and $85\%$ of the total energy transferred to the film for $\Ohf = 0.01$ and $0.1$, respectively. This difference in the film energy distribution hints at the failure of our assumptions to neglect the film's inertia and surface tension. The minimal model is relevant only when the energy transferred to the liquid film is predominantly lost to viscous dissipation. 

Guided by the energy budget analysis in the above two extreme cases, we now evidence the minimal model break down as the film mobility, $\Gamma/\Ohf^{1/3}$, fails to describe the film deflection $\delta_f$. 
In figure~\ref{fig:Deflection}, we report the normalized maximum film deflection $\tilde{\delta}_f = \delta_f/R$ as a function of $\Gamma$ (figure~\ref{fig:Deflection}a) for $\Ohf$ in the range $0.01$ -- $2$ while keeping $\Ohd$ constant.
For $\Ohf > 0.1$, the deflection decreases with increasing $\Ohf$, and the data collapses once the horizontal axis is rescaled by $\Ohf^{-1/3}$ (figure \ref{fig:Deflection}b), confirming the relevance of the film mobility.
However, for lower film Ohnesorge numbers, $\tilde{\delta}_f$ scales with $\Gamma$ independent of $\Ohf$, illustrating the limits of our hypotheses.
Here, one might be tempted to empirically replace the effective control parameter in our model $\propto \Ohf\Gamma^{-3}$ with $\propto \Gamma$, in light of the $\tilde{\delta}_f$ collapse with $\sim \Gamma$ in the low $\Ohf$ regime. However, such a replacement still fails to appropriately account for the kinetic and surface energies of the film. Indeed, low $\Ohf$ films are associated to surface waves, and the maximum deflection $\delta_f$ might not be the correct length scale to mimic their behavior in a simplified model. 
As future work, it would be interesting to couple a linearized quasi-potential fluid model \citep{lee2008impact, galeano2017non, galeano2021capillary} for the liquid pool/film with a spring-mass-damper system for the liquid drop to further investigate this regime. 

\section{Conclusions and Outlook}
\label{sec:Conclusions and Outlook}
\begin{figure}
	\centering
	\includegraphics[width=\textwidth]{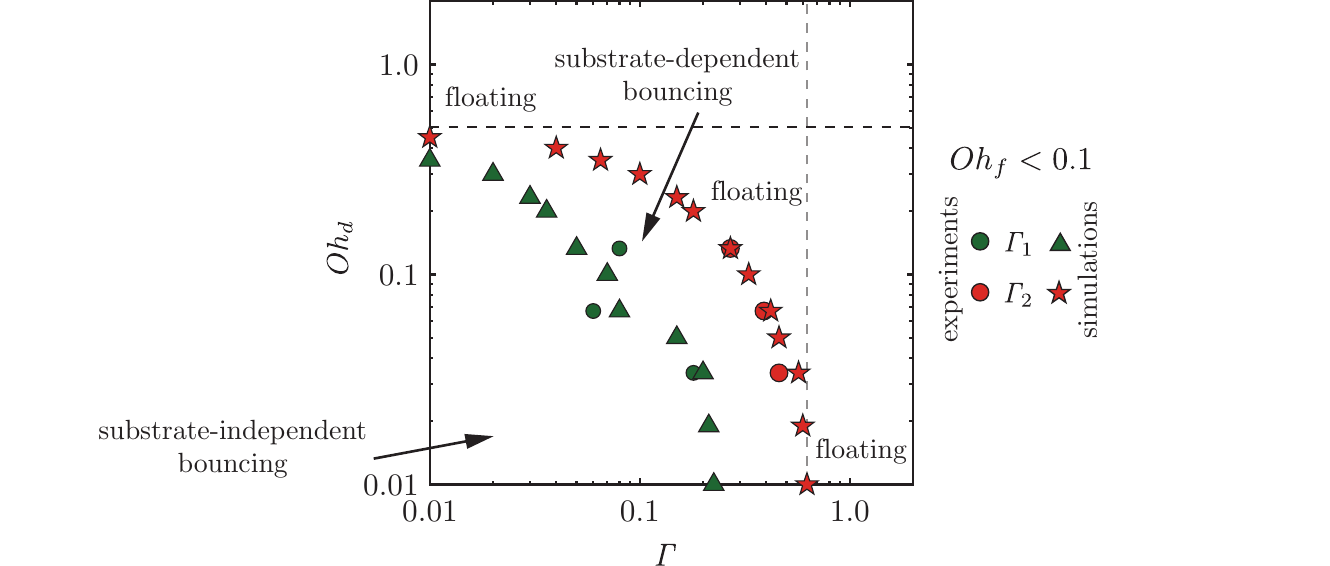}
	\caption{Regime map in terms of the drop Ohnesorge number $\Ohd$ and dimensionless film thickness $\Gamma$ for $\Ohf < 0.1$, showing the transitions between the different regimes identified in this work. $\Gamma_1$ (green symbols) marks the transition from substrate--independent bouncing to substrate--dependent bouncing, whereas $\Gamma_2$ (red symbols) marks the transition from bouncing to non-bouncing (floating). 
		The black dotted line represents the substrate--independent asymptote for the bouncing to non-bouncing (floating) transition (equation~\eqref{eqn:omega1_2}), and the gray dotted line, depicting the inviscid drop asymptote for the bouncing to non-bouncing (floating) transition, is drawn as a guide to the eye.}
	\label{fig:LowOhfRegime}
\end{figure}

In this work, we perform experiments and direct numerical simulations of the rebound of an oil drop impacting on a deformable oil film. We elucidate the role of the drop and film properties: the Ohnesorge numbers of the drop $\Ohd$ and the film $\Ohf$ and the film thickness $\Gamma$ on the impact process.

For films with a low Ohnesorge number ({\it i.e.}, $\Ohf < 0.1$), figure~\ref{fig:LowOhfRegime} summarizes the different regimes identified in this work. For small film thicknesses, we recover the substrate--independent limit  where bouncing is inhibited by the high viscous dissipation in the drop \citep[$Oh_{d,c} \sim \mathcal{O}\left(1\right)$,][]{jha2020viscous}. Increasing the film thickness reduces the drop Ohnesorge number marking the bouncing to non-bouncing (floating) transition as additional energy is transferred to the film, and similarly influences the substrate--independent to substrate--dependent transition. In the inviscid drop limit, bouncing stops once a critical film thickness ($\Gamma_2 \sim \mathcal{O}\left(1\right)$) is reached, independent of $\Ohf$. Here, the invariance of the energy transfer from the drop to the film with $\Ohf$ remains to be explained and deserves further study.


\begin{figure}
	\centering
	\includegraphics[width=\textwidth]{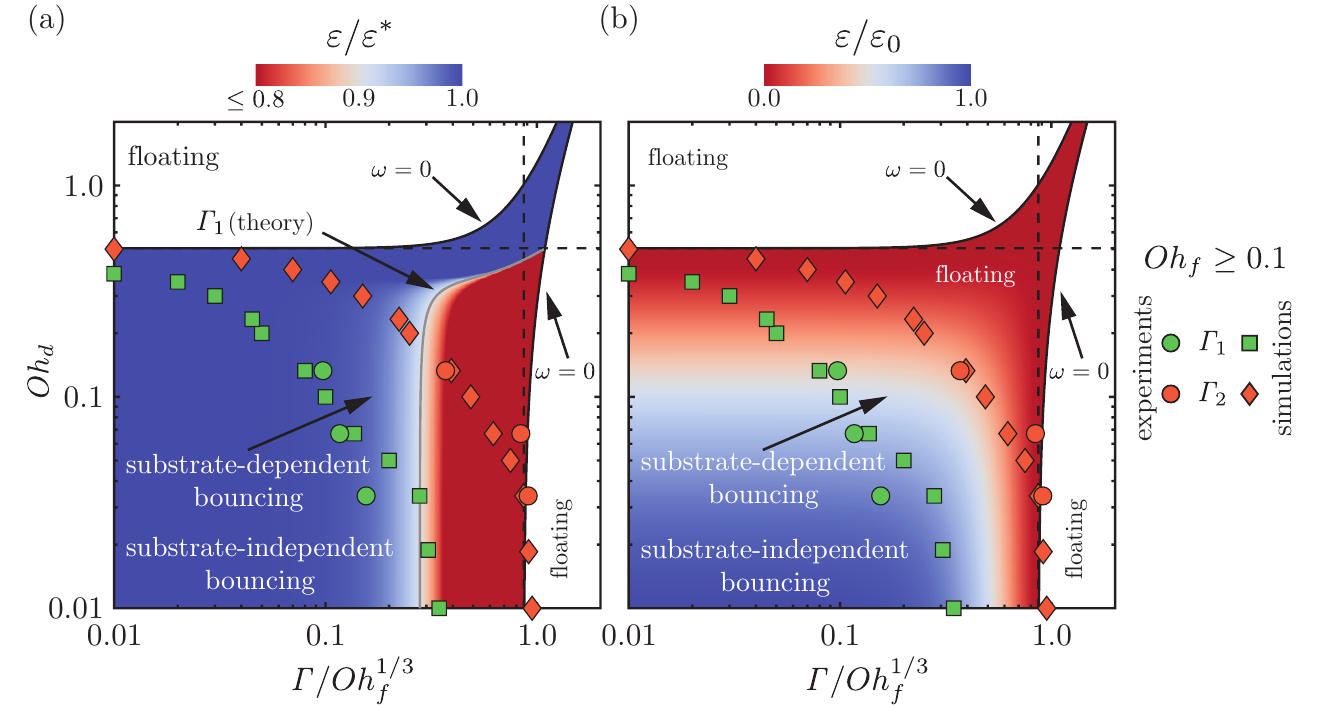}
	\caption{Regime map in terms of the drop Ohnesorge number $\Ohd$ and film mobility $\Gamma/\Ohf^{1/3}$ for $\Ohf \ge 0.1$ showing the transitions between the different regimes identified in this work. 
		$\Gamma_1$ (green symbols) marks the transition from substrate--independent to substrate--dependent bouncing, whereas $\Gamma_2$ (orange symbols) marks the transition from bouncing to non-bouncing (floating).
		The background contour illustrates the theoretical values of the coefficient of restitution $\varepsilon$ (equation~\eqref{eqn:eps_model}) normalized with its (a) substrate--independent limit $\varepsilon^* = \varepsilon\left(\Gamma/\Ohf^{1/3} \to 0\right)$, equation~\eqref{eqn:drySubstrate_epsilon} and (b) inviscid drop and substrate--independent limit $\varepsilon_0 = \varepsilon^*\left(\Ohd \to 0\right)$. The black solid lines shows the predicted bouncing to non-bouncing (floating) transition using the phenomenological model (equations~\eqref{eqn:omega0_2}-\eqref{eqn:omega0_1}), and the black dashed lines show the two asymptotes (equations~\eqref{eqn:omega1_2}-\eqref{eqn:omega1_1}) of bouncing to non-bouncing (floating) regimes. Lastly, in panel (a), the gray solid line shows the prediction for $\Gamma_1$.}
	\label{fig:regimeMap}
\end{figure}

For high Ohnesorge number films ({\it i.e.}, $\Ohf > 0.1$), figure~\ref{fig:regimeMap} summarizes the different regimes identified in this work. Similar to the low $\Ohf$ case, increasing $\Ohd$ and $\Gamma$ inhibits bouncing. In contrast with the previous case, in the inviscid drop limit, the bouncing to non-bouncing (floating) transition occurs at critical film thicknesses that depend on the Ohnesorge number of the film $\left(\Gamma_2 \sim \Ohf^{1/3}\right)$. We propose a minimal phenomenological model describing the key aspects of this process. The background colors in figures~\ref{fig:regimeMap}(a) and~\ref{fig:regimeMap}(b) illustrate the predicted values of the restitution coefficient $\varepsilon$ (equation~\eqref{eqn:eps_model}) normalized with its substrate--independent $\left(\varepsilon^* = \varepsilon\left(\Gamma/\Ohf^{1/3} \to 0\right)\,\text{, equation~\eqref{eqn:drySubstrate_epsilon}}\right)$, and inviscid drop and substrate--independent $\left(\varepsilon_0 = \varepsilon^*\left(\Ohd \to 0\right)\right)$ values, respectively. 
The model accurately predicts the substrate--independent and inviscid drop asymptotes corresponding to the bouncing to non-bouncing (floating) transition, {\it i.e.} $\Gamma_2$.
In the latter limit, the model also captures the substrate--independent to substrate--dependent transition ({\it i.e.}, $\Gamma_1$). Away from these asymptotes, the minimal model fails to predict $\Gamma_1$ and $\Gamma_2$. We attribute this shortcoming to the simplified representation of the drop--film interactions in the model. Nonetheless, notice that the predicted values of the restitution coefficient are very close to zero beyond the bouncing to non-bouncing (floating) transition observed in the simulations. We hypothesize that the model breakdown might be caused by the neglect of gravity which is known to inhibit bouncing \citep{biance2006} and may prevent the take off of drops with small upward velocities. We refer the reader to \citet{vatsalInProgress} for a detailed study of the role of gravity in inhibiting the bouncing of viscous drops.

Finally, we stress that this study does not present an exhaustive exploration of all bouncing regimes. For example, \citet{galeano2021capillary} have shown that spherical hydrophobic solid spheres can bounce off deep low viscosity pools. Consequently, we hypothesize that the bouncing regime could resurrect at high $\Ohd$, $\Gamma$, and low $\Ohf$, evidencing non-monotonic energy transfer. It will be interesting to probe such a regime in future work. \\

\noindent{\bf  Supplementary data\bf{.}} \label{SM} Supplementary material and movies are available at xxxx \\

\noindent{\bf Acknowledgements\bf{.}}  We acknowledge Esra Van't Westende for carrying out initial experiments. We would like to thank Daniel Harris and Andrea Prosperetti for illuminating discussions. This work was carried out on the national e-infrastructure of SURFsara, a subsidiary of SURF cooperation, the collaborative ICT organization for Dutch education and research.\\

\noindent{\bf Funding\bf{.}} We acknowledge the funding by the ERC Advanced Grant No. 740479-DDD, European Union's Horizon 2020 research and innovation programme under the Marie Sklodowska-Curie grant agreement No 722497 and by the Max Planck Center Twente for Complex Fluid Dynamics.\\

\noindent{\bf Declaration of Interests\bf{.}} The authors report no conflict of interest.\\

\noindent{\bf  Author ORCID\bf{.}}  

V. Sanjay \href{https://orcid.org/0000-0002-4293-6099}{https://orcid.org/0000-0002-4293-6099}; 

S. Lakshman \href{https://orcid.org/0000-0002-0563-7022}{https://orcid.org/0000-0002-0563-7022}; 

P. Chantelot \href{https://orcid.org/0000-0003-1342-2539}{https://orcid.org/0000-0003-1342-2539}; 

J. H. Snoeijer \href{https://orcid.org/0000-0001-6842-3024}{https://orcid.org/0000-0001-6842-3024}; 

D. Lohse \href{https://orcid.org/0000-0003-4138-2255}{https://orcid.org/0000-0003-4138-2255}. \\


\appendix
\section{Air layer rupture}
\label{App:FilmRupture}
\begin{figure}
	\centering
	\includegraphics[width=\textwidth]{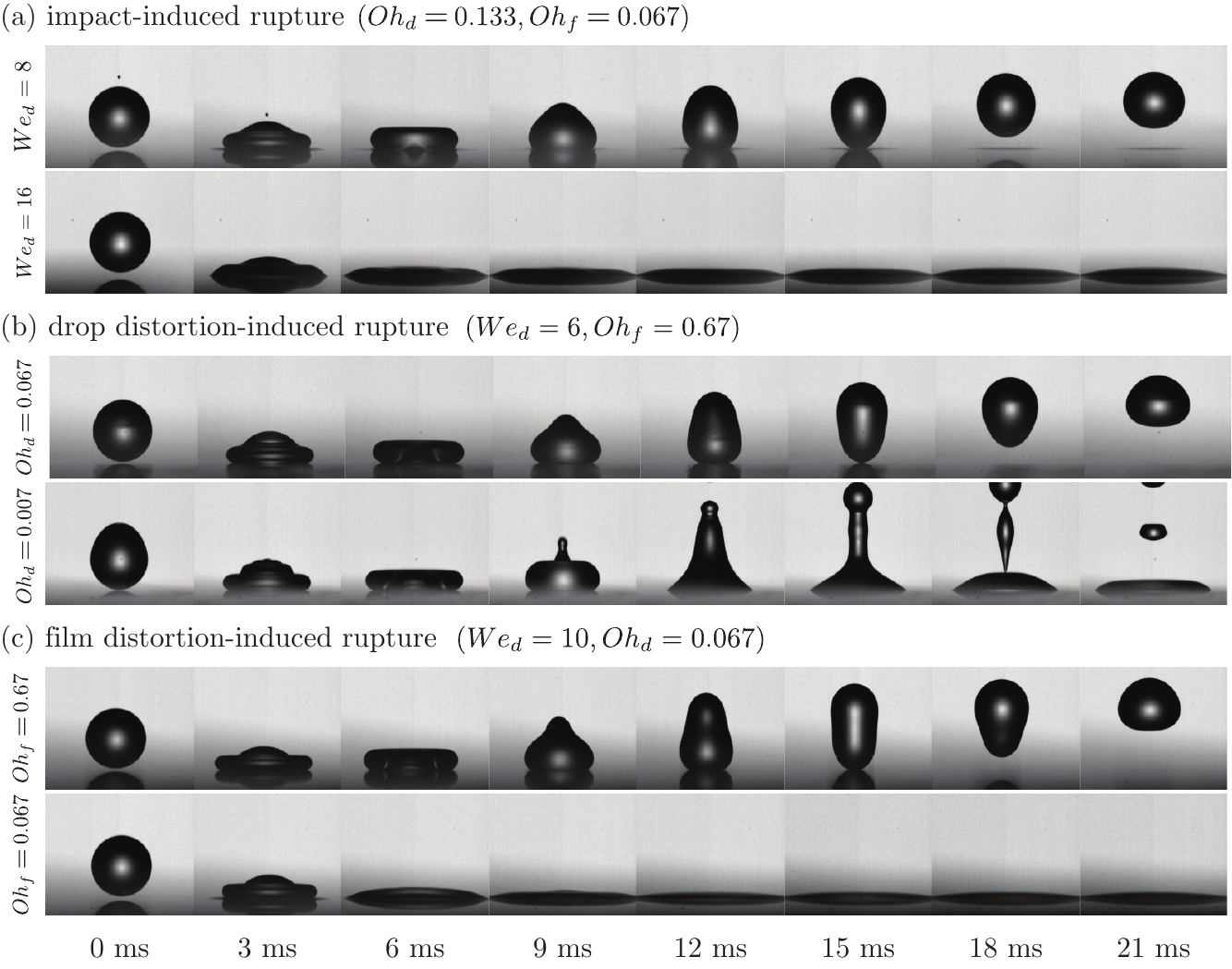}
	\caption{Rupture of the air layer and subsequent coalescence of impacting drops with the liquid coating. 
		Rupture can occur due to: (a) impact as $\Wen$ increases, (b) downward jetting as $\Ohd$ decreases, and (c) film distortions due to surface waves as $\Ohf$ decreases. 
		For panels (a) and (c), $\Gamma = 0.03$, and for panel (b) $\Gamma = 0.01$ (dry-substrate limit).}
	\label{fig:figureA1}
\end{figure}

%
\revRev{We investigate drop bouncing off viscous liquid films that mimic atomically smooth substrates. The occurrence of such rebounds is tied to the existence of a stable air layer, enabling drop levitation. It is thus important to determine the conditions leading to air film rupture in terms of our control parameters.}
Figure \ref{fig:figureA1}(a) illustrates the air layer break up at large Weber numbers. The air film fails during drop spreading as the intervening air layer drains below a critical thickness on the order of $10 - 100\,\si{\nano\meter}$, characteristic of the range of van der Waals forces \citep{charles1960coalescence, SprittlesPhysRevLett.124.084501, zhang2021thin}.
Figure \ref{fig:figureA1}(b) evidences the influence of the drop Ohnesorge number $\Ohd$ on the coalescence transition. 
At low $\Ohd$, the convergence of capillary waves at the drop apex, during the retraction phase, can create an upward Worthington jet and an associated downward jet due to momentum conservation \citep{Bartolo2006Singular,lee2020downward,zhang2022impact}. 
This downward jet can puncture the air film and lead to coalescence during the drop retraction.
Lastly, the air layer can also break due to surface waves on low $\Ohf$ films (see figure~\ref{fig:figureA1}c). 

In summary, figure~\ref{fig:figureA1} shows that the critical Weber number beyond which the air layer between the drop and the film ruptures is sensitive to the Ohnesorge numbers of both the drop and the film \citep{tang2016nonmonotonic, tang2018bouncing}, and that the bouncing to coalescence transition can arrest the superamphiphobic-type rebounds discussed in this work.
\revRev{For completeness, we also mention that a second coalescence transition occurs in our experiments, at times at least one order of magnitude larger than that associated to drop rebound, when the air film trapped below a floating drop drains \citep{lo2017mechanism,duchemin2020dimple}.}
The analysis of \revRev{both these} transitions is beyond the scope of the present study and we refer the reader to \citet{lohse-2020-pnas, SprittlesPhysRevLett.124.084501} for further discussion and review on this topic.

\section{Substrate independent bouncing}
\label{app:SubstrateIndependentBouncing}

As the film thickness decreases or the film viscosity increases, the impact process becomes independent of the film properties. 
In this limit, $\Gamma/\Ohf^{1/3} \to 0$, the phenomenological model predictions for the contact time and restitution coefficient, equations~\eqref{eqn:tc_model} and \eqref{eqn:eps_model}, become

\revRev{\begin{align}
		\label{eqn:drySubstrate_time}
		t_c^*(\Ohd) =\,&t_c(\Ohd, \Gamma/\Ohf^{1/3} \to 0) = \tau_\gamma\left(\frac{2\pi}{\sqrt{4c_{k} - c_{d}^{2} \Ohd^{2}}}\right),\\
		\label{eqn:drySubstrate_epsilon}
		\varepsilon^*(\Ohd, \Wen) =\,&\varepsilon(\Ohd, \Wen, \Gamma/\Ohf^{1/3} \to 0) = \varepsilon_0(\Wen)\exp\left( \frac{-\pi c_{d} \Ohd}{\sqrt{4c_{k} - c_{d}^{2} \Ohd^{2}}} \right),
\end{align}}

\noindent which are identical to the predictions obtained by \citet{jha2020viscous} for the impact of viscous drops on a superhydrophobic surface. 

\begin{figure}
	\centering
	\includegraphics[width=\textwidth]{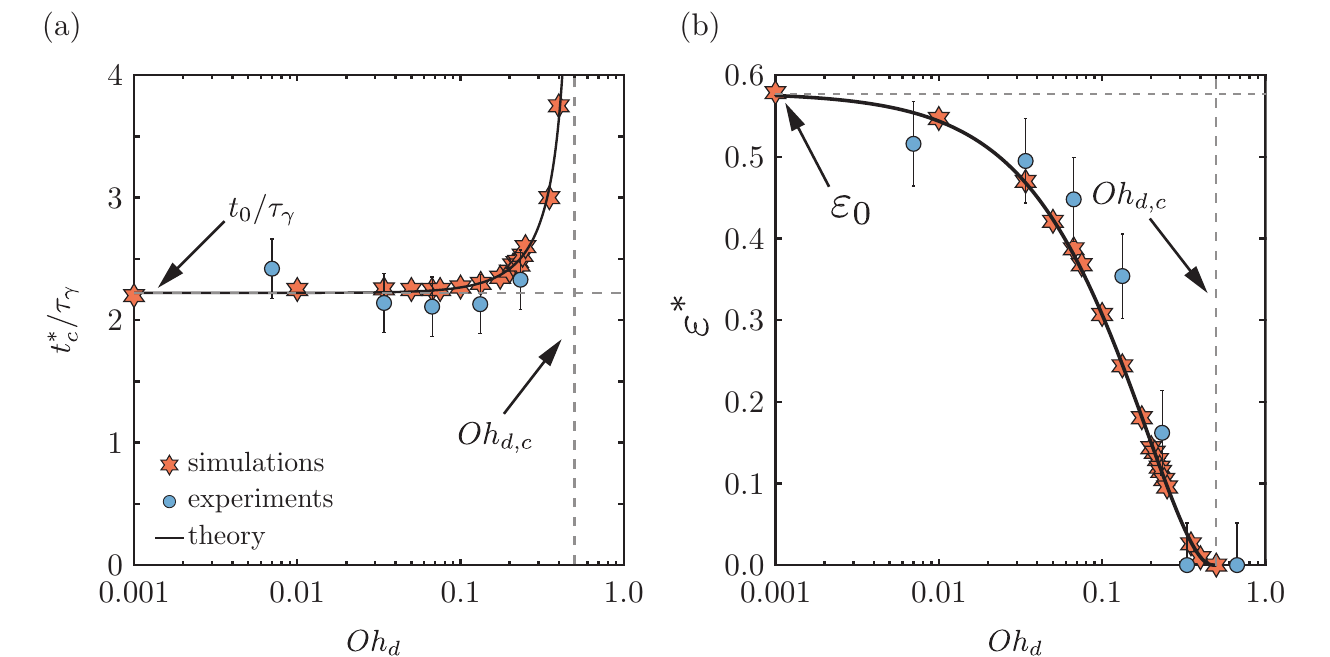}
	\caption{Substrate independent bouncing: variation of (a) the contact time $t_c^*$ normalized with the inertio-capillary timescale $\tau_\gamma$, and (b) the coefficient of restitution $\varepsilon^*$ with the drop Ohnesorge number $\Ohd$. The solid black lines represent equations~\eqref{eqn:drySubstrate_time}-\eqref{eqn:drySubstrate_epsilon}. These predictions are consistent with \revRev{those} of \citet{jha2020viscous}, and set the prefactors $c_{k}$ and $c_{d}$ to $2.0 \pm 0.1$ and $5.6 \pm 0.1$, respectively. Here, $\Wen = 4$ and $Bo = 0.5$.}
	\label{fig:appendix_drylimit_figure}
\end{figure}

Reducing equation~\eqref{eqn:drySubstrate_time} to the case of low viscosity drops ($\Ohd \to 0$), we get $t_0/\tau_\gamma = \pi/\sqrt{c_k}$, as expected from the water-spring analogy \citep{richard2002contact, okumura2003water}.
We thus determine the prefactor $c_k$ by fitting the inviscid limit of our data, $t_0 = 2.2\tau_\gamma$ (figure~\ref{fig:appendix_drylimit_figure}a), yielding $c_k = \left(\pi\tau_\gamma/t_0\right)^2 \approx 2$.

Furthermore, applying a least square fit to our experimental and numerical data for the coefficient of restitution, which decays exponentially with increasing $\Ohd$ (figure~\ref{fig:appendix_drylimit_figure}b), allows us to fix $c_d = 5.6 \pm 0.1$. Lastly, the model predicts the existence of a critical Ohnesorge number $Oh_{d,c} = 2\sqrt{c_k}/c_d \approx 0.5$ above which the drops do not bounce. This asymptote is in quantitative agreement with our data (see the dashed gray lines in figure~\ref{fig:appendix_drylimit_figure}).  

Finally, we compare the above value of $c_d$ to that obtained by \citet{jha2020viscous}.
To do so, we note that \citet{jha2020viscous} further reduced equation~\eqref{eqn:drySubstrate_epsilon} to \revRev{$\varepsilon^*(\Wen, \Ohd) \approx \varepsilon_0(\Wen)\exp\left(-\alpha\Ohd\right)$}, where $\alpha = 2.5 \pm 0.5$ fits their experimental data, independent of the impact Weber number. 
The equivalent fitting parameter in our case is $\alpha = (\pi/2) c_d/\sqrt{c_k} \approx 6$. This discrepancy can be attributed to the different values of the critical Ohnesorge number $Oh_{d,c}$  which could stem from the Bond number variation between the two cases: $(Oh_{d,c}, Bo) \approx (0.8, 0.2)$, in \citet{jha2020viscous}, and $(0.5, 0.5)$ in this work. 
Exploring the influence of $Bo$ is beyond the scope of this work and we refer the reader to \citet{vatsalInProgress} for detailed discussions.

\section{Influence of the impact Weber number}
\label{sec:weber_influence}

\begin{figure}
	\centering
	\includegraphics[width=\textwidth]{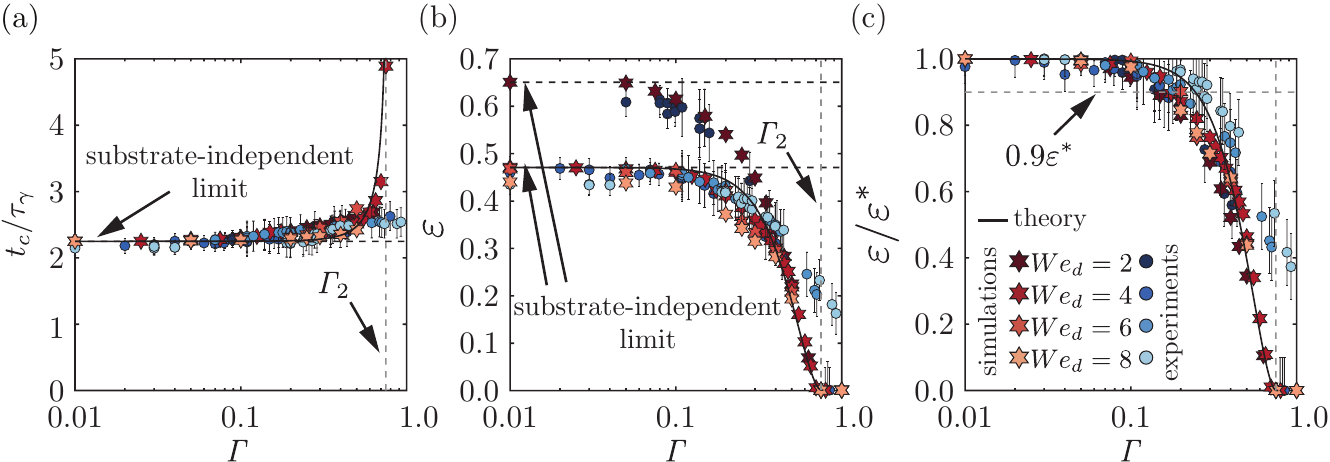}
	\caption{Influence of the impact Weber number on the rebound: variation of (a) the contact time $t_c$ normalized with the inertio-capillary time scale $\tau_\gamma$, (b) the restitution coefficient, and (c) the restitution coefficient normalized with its dry substrate value as a function of the dimensionless film thickness $\Gamma$. 
		Here, ($\Ohd, \Ohf$) = ($0.034, 0.67$). 
		In each panel, solid black lines represent the model prediction for ($c_k, c_d, c_f$) = ($2, 5.6, 0.46$) and the vertical dashed gray line indicate $\Gamma_2$, above which drops do not bounce. 
		In panels (a) and (b), black dashed lines show the \revRev{$\Wen$-dependent values in the dry substrate limit at fixed $\Ohd$, i.e., $\varepsilon^*(\Ohd = 0.034, \Wen)$.}
		Lastly, in panel (c), the horizontal dashed gray line denotes the $0.9\varepsilon^*$ criterion used to determine the substrate--independent to substrate--dependent transition for bouncing drops.}
	\label{fig:appendix_WeberVariation}
\end{figure}

Figure~\ref{fig:appendix_WeberVariation} describes the influence of the Weber number $\Wen$ on the drop impact process for a representative case with $\Ohd = 0.034$ and $\Ohf = 0.67$.
Both the contact time (figure~\ref{fig:appendix_WeberVariation}a) and the coefficient of restitution (figure~\ref{fig:appendix_WeberVariation}b) are fairly independent of the Weber number for $\Wen \ge 4$. 
Furthermore, normalizing $\varepsilon$ with its $\Wen$-dependent value in the dry substrate limit \revRev{$\varepsilon^*(\Ohd, \Wen)$}, at fixed $\Ohd$ \revRev{($= 0.034$ in figure~\ref{fig:appendix_WeberVariation})}, we observe a collapse for $\Wen = 2 - 8$, similar to that obtained by \citet{jha2020viscous}. 
Readers are referred to \citet{vatsalInProgress} for detailed discussions on the mechanisms of the influence of the Weber number on the coefficient of restitution. 

\section{Measuring the film thickness}
\label{sec:measuring_film_thickness}
\begin{figure}
	\centering
	\includegraphics[width=\textwidth]{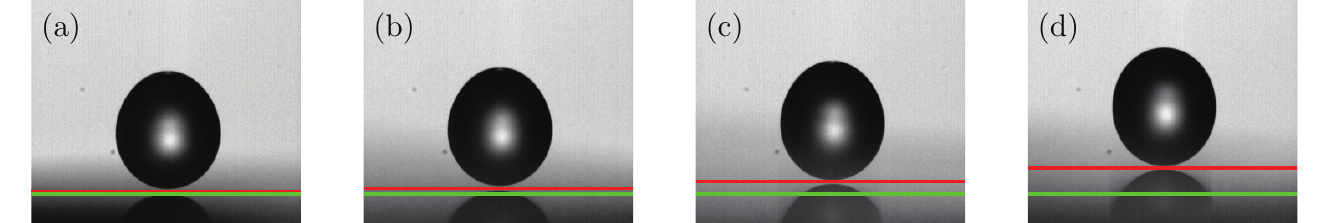}
	\caption{Experimental side view snapshots at the instant of impact, $t = 0$. Each snapshot shows the wall location denoted by a green horizontal line and the free-film interface denoted by a red horizontal line. 
		The film thickness is estimated from the vertical difference between the two lines which results in dimensionless film thickness of $\Gamma = h_{f}/R =0.05$ (a), $0.11$ (b), $0.23$ (c), and $0.48$ (d).}
	\label{fig:figureA2a}
\end{figure}

Silicone oil films with thicknesses $h_f < 30\,\si{\micro\meter}$, are prepared using spin coating and measured using reflectometry \citep{reizman1965optical}. 
Thicker films ($h_f > 30\,\si{\micro\meter}$) are prepared by depositing a controlled volume of silicone oil on a glass slide. 
The film thickness is then measured using side view imaging by locating the vertical position of the glass slide wall (green line in figure \ref{fig:figureA2a}) and of the film free surface (red line in figure \ref{fig:figureA2a}).
The uncertainty in the film thickness measurement using this method is about $\pm 30\,\si{\micro\meter}$, which corresponds to an uncertainty of about 3 pixels.

\section{Measuring the restitution coefficient}
\label{sec:restitution in simulations}

In this appendix, we describe the procedure used to determine the restitution coefficient. 
In experiments, we measure the drop's maximum center of mass height relative to the undisturbed film surface to get the restitution coefficient as $\varepsilon = \sqrt{2g(H-R)}/V$, where $V$ is the impact velocity.  
In simulations, we measure the coefficient of restitution as the ratio of the take-off velocity $v_{\text{cm}}(t_c)$ to the impact velocity $V$,

\begin{align}
	\varepsilon = \frac{v_{\text{cm}}(t_c)}{V}
\end{align}

\noindent where $t_c$ denotes the contact time. 
The latter definition requires to precisely evaluate the contact time $t_c$.
This is difficult as a thin film of air is always present between the drop and the film surface, we assume ideal non-coalescence between the drop and the film.
In simulations, we automatise the detection of the end of apparent contact by taking $t_c$ as the instant when the normal reaction force $F(t)$ between the film and the drop is zero \citep[for details on the force calculation, see][]{zhang2022impact}.
If the center of mass velocity $v_{\text{cm}}(t_c)$ is not in the upward direction ({\it i.e.} it is zero or negative), we categorize the case as non-bouncing.

\section{Code availability}
The codes used in the present article are permanently available at \citet{basiliskVatsalDropFilm}.

\bibliographystyle{jfm}
\bibliography{BouncingDrops}

\end{document}